\documentclass[reprint, amsmath,amssymb, aps,pra,
 floatfix, nobalancelastpage
]{revtex4-2}

\usepackage[caption=false]{subfig} % works well with revtex4-2
\usepackage{bm}
\usepackage[dvipsnames]{xcolor}
\usepackage[pdftex]{graphicx}% Include figure files
\usepackage{dcolumn}% Align table columns on decimal point
\usepackage{bm}% bold math
\usepackage{tikz}
\usepackage{amsthm}
\usepackage[linesnumbered,ruled,vlined]{algorithm2e}  % For algorithms
\usepackage{algpseudocode}  % For pseudocode
\usepackage{graphicx}
\usepackage{quantikz}
\usepackage[table]{xcolor}
\usepackage{colortbl} % optional
\usepackage{graphicx} % for \resizebox
\usepackage{mathrsfs}

\usepackage[normalem]{ulem}

\usepackage{booktabs}
\usepackage{pgfplots}
\pgfplotsset{compat=1.18}
\providecommand{\thresholdQDataPath}{.}

\usetikzlibrary{fit,backgrounds}

\definecolor{conflictRed}{HTML}{A84A4A}
\definecolor{choiceBlue}{HTML}{176B87}
\definecolor{choiceFill}{HTML}{DCEFF4}
\definecolor{partFill}{HTML}{F3F5F7}
\definecolor{neutralGray}{HTML}{68727D}
\usetikzlibrary{arrows.meta,positioning,calc}

\definecolor{searchBlue}{HTML}{176B86}
\definecolor{searchFill}{HTML}{E5F1F5}
\definecolor{witnessGreen}{HTML}{2F7653}
\definecolor{witnessFill}{HTML}{E6F2EA}
\definecolor{nonbareRed}{HTML}{A34A4A}
\definecolor{nonbareFill}{HTML}{F7E7E7}
\definecolor{neutralGray}{HTML}{66717C}

\usepackage{microtype}

\makeatletter
\expandafter\let\csname c@theorem\endcsname\relax
\expandafter\let\csname c@lemma\endcsname\relax
\expandafter\let\csname c@proposition\endcsname\relax
\expandafter\let\csname c@corollary\endcsname\relax
\expandafter\let\csname c@definition\endcsname\relax
\expandafter\let\csname c@remark\endcsname\relax
\makeatother

\newtheorem{theorem}{Theorem}
\newtheorem{lemma}[theorem]{Lemma}
\newtheorem{proposition}[theorem]{Proposition}
\newtheorem{corollary}[theorem]{Corollary}

\theoremstyle{definition}
\newtheorem{definition}[theorem]{Definition}

\theoremstyle{remark}

\newcommand{\cP}{\mathcal{P}}
\newcommand{\cS}{\mathcal{S}}
\newcommand{\cN}{\mathcal{N}}

\newcommand{\Ftwo}{\mathbb{F}_2}

\SetKwInput{KwInput}{Input} 
\SetKwInput{KwOutput}{Output}  
\SetKwInput{KwInit}{Initialization}   
\SetKwInput{KwStep}{Step1}
\SetKwInput{KwStepp}{Step2}
\SetKwInput{KwSteppp}{Step3}

\newcommand{\PadCell}[1]{\kern8pt#1\kern8pt}
\newcommand{\PadCellColor}[2]{\cellcolor{#1}\kern8pt#2\kern8pt}

\usepackage{hyperref}% add hypertext capabilities
\hypersetup{
  colorlinks   = true, % Colours links instead of ugly boxes
  urlcolor     = blue, % Colour for external hyperlinks
  linkcolor    = blue, % Colour of internal links
  citecolor    = blue  % Colour of citations
}

\begin{document}

\preprint{APS/123-QED}

\title{\color{black}Bare-ancilla fault-tolerant syndrome extraction: General extensions of distance-three codes and structural criteria for graph codes}

\author{Harsh Gupta}
\email{harsh22@iiserb.ac.in}
 \affiliation{
 Department of Electrical Engineering and Computer Science, Indian Institute of Science Education and Research Bhopal, 462066 Bhopal, India }
\author{Mainak Bhattacharyya}
 \email{mainak23@iiserb.ac.in}
\affiliation{
 Department of Electrical Engineering and Computer Science, Indian Institute of Science Education and Research Bhopal, 462066 Bhopal, India }
\author{Ritik Jain}
 \email{ritikjain@iiserb.ac.in}
\affiliation{
 Department of Electrical Engineering and Computer Science, Indian Institute of Science Education and Research Bhopal, 462066 Bhopal, India }
\author{Ankur Raina}
 \email{corresponding author: ankur@iiserb.ac.in}
\affiliation{
 Department of Electrical Engineering and Computer Science, Indian Institute of Science     Education and Research Bhopal, 462066 Bhopal, India }

\date{\today}

\begin{abstract}

\noindent The reliability of quantum computation critically depends on the performance of quantum error-correcting codes (QECCs).
Performance of QECCs can be severely degraded by hook errors, which effectively reduce the code distance.
\textcolor{black}{
We develop a systematic framework for bare-ancilla syndrome extraction in $[[n,k,3]]$ stabilizer codes.
It provides a criterion for a specific single ancilla fault model and jointly searches for suitable stabilizer generators and data-ancilla interaction orders to correct hook errors. 
We call such stabilizer generator sets bare-ancilla codes (BACs). 
We also present two constructions that enlarge the set of valid BACs while preserving minimum distance, either by keeping the number of logical qubits fixed or by increasing it.
Under the correlated-data error model, $3177$ $[[7,1,3]]$ graph codes are found to be BACs out of $3379$ and all $335415$ $[[8,1,3]]$ graph codes are BACs.
We tested $500$ $[[7,1,3]]$ and $1000$ $[[8,1,3]]$ graph codes in the presence of anisotropic and depolarizing noise.
The number of data-ancilla interactions is an important predictor of the pseudo-threshold of these graph codes.
In the analyzed codes, the BAC, which works with the bare ancilla method, usually performs well or better than the flag method, particularly for depolarizing noise.
Notably, we report  new bare ancilla codes, namely $[[6,1,3]]$  and $[[7,1,3]]$  with improved code rate compared to the bare code used in the work of Muyuan Li \emph{et al.} and Maheshwari \emph{et al.} respectively.}

\end{abstract}

\maketitle

\section{Introduction}
\label{sec:introduction}

% \noindent The evolution of hardware architectures for quantum computing has been rapid in recent years to support a reliable fault-tolerant framework for the quantum protocols.
% Manipulation of the logical information encoded on the physical qubits should yield high fidelity, indicating highly error-suppressed outcomes.
% As a result, the development of a robust Quantum Error Correcting Code (QECC) is fundamental in the advancement of fault-tolerant (FT) quantum computing.
% Fault-tolerant quantum computation can suppress logical errors when the physical error rate is below a suitable threshold ~\cite{10.1145/258533.258579,nielsen2010quantum}.
% However, encoding information in a quantum code is not sufficient by itself because during syndrome measurement using the stabilizer, an ancilla qubit interacts with several data qubits.
% A fault on this ancilla propagates through the remaining two-qubit gates, creating a correlated data error, which needs to be corrected.
% The propagated error depends on the measured stabilizer, the order of the data-ancilla interactions, the fault location, and the assumed fault model. 
% Fault tolerance is therefore a property of both the code and its syndrome-extraction circuit.
\noindent The evolution of hardware architectures for quantum computing has been rapid in recent years to support a reliable fault-tolerant framework for the quantum protocols.
Manipulation of the logical information encoded on the physical qubits should yield high fidelity, indicating highly error-suppressed outcomes.
As a result, the development of a robust Quantum Error Correcting Code (QECC) is fundamental in the advancement of fault-tolerant (FT) quantum computing.
Implementing QECCs in practical scenarios poses significant challenges, particularly due to physical errors that affect the error detection process. 
During a computation, fault tolerance can be achieved asymptotically if the physical error rate is below a certain threshold~\cite{10.1145/258533.258579, nielsen2010quantum}.
Crucially, the syndrome extraction process must also be implemented in an FT manner to prevent errors from propagating into the data qubits.
One of the earliest codes tested for fault tolerance on small systems is the  $[[4,2,2]]$ QECC using ion trap and superconducting qubits \cite{10.5555/3370220.3370224, PhysRevLett.119.180501}. They demonstrated that error detection can improve certain simple sampling tasks. 
This code only detects a Pauli error as its minimum distance is two.
Any limited error-correcting capacity of a QECC imposes significant resource overhead required for FT quantum computation.
Several techniques have been proposed for extracting FT syndrome.
Steane's method requires an FT logical state in the syndrome measurement, whereas Shor's method requires extra qubits prepared in the cat state for FT syndrome measurement~\cite{PhysRevLett.78.2252,shor1996}.
Similarly, Knill's method uses a logical Bell pair~\cite{Knill2005}.

\noindent Researchers have shown that stabilizers outside the minimal generating set can be leveraged to strengthen the error-correcting capability of quantum codes~\cite{fujiwara2015high}.
Fujiwara proved that it is possible to detect the measurement error of the syndrome qubit even in $[[5, 1, 3]]$ QECC~\cite{fujiwara2014ability}.
Yoder \emph{et al.} address that both $[[5,1,3]]$ and $[[7,1,3]]$ QECCs can be made resistant to hook errors using three and two additional qubits, respectively~\cite{Yoder2017surfacecodetwist}. 
Using an extra two qubits, Chao and Reichardt were able to correct the faults of distance-three codes that lead to correlated errors in data qubits during syndrome extraction~\cite{PhysRevLett.121.050502}.
Muyuan Li \emph{et al.} proposed the bare method, a FT syndrome-extraction scheme for a $[[7,1,3]]$ QECC that uses only one ancillary qubit by cleverly reordering the stabilizer's operators.
This method ensures no logical error occurs in data qubits due to a single Pauli error on the ancillary qubit~\cite{Muyuna}.
The authors in Ref. \cite{Muyuna} find the logical error rates for anisotropic noise below the pseudo-threshold limit. 
\textcolor{black}{Maheshwari\emph{et al.} later studied a selected $[[8,1,3]]$ non-CSS code under anisotropic and depolarizing noise ~\cite{maheshwari2024fault}.
These examples show that carefully ordered bare-ancilla circuits can reduce ancillary-qubit overhead.
However, they do not provide a general method that jointly searches the stabilizer generating set and all allowed interaction orders}

% \noindent Several methods have been developed to make syndrome extraction FT.
% Steane's method uses an encoded ancilla state, Shor's method uses a verified cat state, and Knill's method uses an encoded Bell pair~\cite{PhysRevLett.78.2252,shor1996,Knill2005}.
% Flag-based methods reduce this overhead by using a syndrome ancilla together with one or more flag qubits to detect faults in ancilla ~\cite{PhysRevLett.121.050502,chamberland2017flag,chao2020flag}.
% These methods apply to broad classes of stabilizer codes, but still require additional ancillary resources.
% The bare-ancilla method uses only a syndrome ancilla and controls hook errors through the stabilizer generators and their interaction orders.
% Muyuan Li\emph{et al.} presented a $[[7,1,3]$ non-CSS code for which a suitable interaction order makes the propagated error correctable under the anisotropic noise model considered in their work ~\cite{Muyuna}.
% Maheshwari\emph{et al.} later studied a selected $[[8,1,3]]$ non-CSS code under anisotropic and depolarizing noise ~\cite{maheshwari2024fault}.
% These examples show that carefully ordered bare-ancilla circuits can reduce ancillary-qubit overhead.
% However, they do not provide a general method that jointly searches the stabilizer generating set and all allowed interaction orders.

\color{black}
\noindent This work addresses that problem for distance-three stabilizer codes.
A bare-ancilla code(BAC) is specified by ($\mathcal C, \Pi, \mathcal M$) where $\mathcal C$ is the stabilizer code, $\Pi$ specifies the measured stabilizer generators interaction orders, and $\mathcal M$ is the single-fault model.
Thus, the term BAC does not refer to $\mathcal C$ alone.
Two algorithms are used to find a valid BAC.
First, used to find the valid permutation of operators for each generator, and secondly, to perform a global search for the interaction orders and stabilizer generators of a given code.

\noindent For a valid $[[n,k,3]]$ BAC we give the two constructions which extend it to $[[n',k',3]]$ BAC with $n'>n$ and $k'>k$ while keeping the minimum distance constant at 3. 
First, the pendent extension adds the physical qubits while preserving the logical information, minimum distance, and bare-ancilla condition under the stated compatibility assumptions. 
Secondly, blockwise composition combines the independent BACs and increases the logical information as the number of encoded qubits increases.
Together, these constructions provide BACs with any chosen number of logical qubits and any sufficiently large number of physical qubits, provided suitable parent BACs exist under the same fault model.

\noindent Graph codes ~\cite{Schlingemann_2001,PhysRevA.69.022316,cafaro:hal-02287093,schlingemann2001stabilizercodesrealizedgraph} provide a useful setting in which the conditions for selecting valid stabilizer generators and interaction orders can be expressed directly in terms of the parent adjacency matrix and the message neighborhood vector.
Graph code search shows that the stabilizer generator set is an essential part of BACs and how hook errors depend on various graph parameters and other properties.  
Under the correlated ancilla-data fault model, $3177$ of the $3379$ $[[7,1,3]]$ stabilizer codes are valid BACs, while $202$ are non-BACs.
However, all the $335415$ tested $[[8,1,3]]$ graph codes are BACs. 
\textcolor{black}{We also obtained  $[[6,1,3]]$ and $[[7,1,3]]$ codes for the ancilla-only and correlated ancilla-data fault models, respectively}.
These are the optimized codes in terms of code rate for their respective fault models, to the best of our knowledge.
Finally, circuit-level simulation evaluate $500$ $[[7,1,3]]$ and $1000$ $[[8,1,3]]$ BACs under the anisotropic and depolarizing noise.
The effective number of data-ancilla interactions, denoted by $Q$, is closely related to the observed pseudo-threshold.
\textcolor{black}{For the simulated codes, the BAC, which works with the bare ancilla method,  usually matches or outperforms the corresponding flag method, with the clearest advantage under depolarizing noise.}

\noindent The remainder of the paper is organized as follows. 
Stabilizer codes, hook errors, and fault-tolerant syndrome extraction are reviewed in Section~\ref{sec:Preliminaries}.
The exact bare-ancilla condition and an algorithm for finding a valid stabilizer set using stabilizer generator-and interaction-order search are presented in Section~\ref{sec: Bare construction}.
Section~\ref{sec:general-nk3-extensions} then develops the pendent extension and blockwise composition.
The graph-message representation along with graph code construction is introduced in Section~\ref{sec: MBQC construct QECC}, followed by the algebraic conditions of graph codes on hook errors and $[[7,1,3]]$ and $[[8,1,3]]$ graph code classification in Section~\ref{sec:graph-theoretic-bac}.
The simulation protocol and logical performance analysis are discussed in Section~\ref{sec:numerical-graph-performance}, and the paper concludes in Section~\ref{sec:conclusion}.
\color{black}

\section{Background and Preliminaries}
\label{sec:Preliminaries}
\noindent QECCs form the foundation for protecting quantum information against decoherence and operational errors.
In this section, we shall briefly discuss some basic concepts necessary to understand the rest of the paper.

\subsection{CSS and non-CSS codes}

\noindent In 1995, Shor introduced the first scheme capable of correcting arbitrary single-qubit errors using nine physical qubits, achieving a code rate of $1/9$~\cite{shor9qubit1995}.
Steane subsequently introduced a seven-qubit code with a higher code rate of $1/7$~\cite{steane1996}.
Laflamme \emph{et al.} later established the five-qubit ``perfect'' code, which saturates the quantum Hamming bound and achieves the minimum block size required to protect a single logical qubit from any single-qubit error~\cite{laflamme1996perfect}.
While these codes differ in code length, they share the same minimum distance and therefore correct arbitrary single-qubit errors.
These examples are described uniformly by the stabilizer formalism~\cite{gottesman1997stabilizercodesquantumerror}.

\noindent An $[[n,k,d]]$ stabilizer code is the joint $+1$ eigenspace of $n-k$ independent mutually commuting Pauli operators, with $-I$ excluded from the group generated by those operators.
Let $\mathcal P_n$ be the $ n$-qubit Pauli group up to an overall phase, and $\mathcal S\subseteq\mathcal P_n$ be the corresponding stabilizer group.
A CSS code has a generating set whose Pauli operators are purely $X$-type or purely
$Z$-type~\cite{Calderbank_1996}.
If $C_X=\ker H_X$ and $C_Z=\ker H_Z$, the CSS commutation condition is $C_Z^\perp\subseteq C_X$, equivalently $C_X^\perp\subseteq C_Z$.
A non-CSS stabilizer code, the stabilizer set must have some generator with mixed $X$, $Y$, and $Z$ Pauli operators~\cite{laflamme1996perfect,leverrier2022quantum,srivastava2022xyz}.
For a parity check matrix $H=[H_X\mid H_Z]$ over $\mathbb F_2$, mutual commutativity of their corresponding stabilizer generators is equivalent to
\begin{align}
 H_Z H_X^T+ \ H_X H_Z^T = \mathbf{0} \pmod 2. \nonumber
\end{align}
However, satisfying the stabilizer commutation relations alone is insufficient for achieving fault tolerance during syndrome extraction.
In practical implementations, faults on ancillary qubits may propagate via two-qubit interactions, generating correlated errors on the data qubits, commonly known as hook errors.

\subsection{Hook Errors in QEC}

\noindent Syndrome measurements in quantum error correction are typically performed by coupling ancilla qubits to several data qubits through two-qubit controlled gates such as $CNOT$ or $CZ$ gates~\cite{nielsen2010quantum}.
Pauli errors on the ancilla qubit can propagate through these interactions, generating errors on the data qubits.
They are called hook errors, since a single ancilla fault can effectively hook onto multiple qubits instead of remaining confined to one~\cite{10.1063/1.1499754}. 
Consequently, a fault on the ancilla qubit can propagate through later two-qubit gates between data-ancilla qubits and create a correlated error on two or more data qubits.
Such propagation depends on the circuit, the fault location, and the type of Pauli error that occurs on the ancilla.
If untreated, hook errors reduce the effective code distance, thereby significantly limiting the code's error-correcting capability.
Hence, FT syndrome extraction requires careful circuit design that mitigates hook errors by ensuring that the resulting errors remain correctable.
This requirement motivates the FT syndrome-extraction schemes reviewed next.

\subsection{Fault-tolerant syndrome-extraction schemes}

\noindent Over the past decade, major advances have been made in developing fault-tolerant schemes for syndrome measurement.
Fujiwara showed that suitable redundant stabilizer measurements can protect against faults in the extracted syndrome ~\cite{fujiwara2014ability}.
Building on this idea, Ashikhmin, Lai, and Brun subsequently used classical coding of syndrome bits to reduce repeated-measurement overhead and improve robustness against syndrome measurement errors~\cite{ashikhmin2014robust}.
Crow \emph{et al.} developed a coherent, measurement-free error-correction scheme and analyzed its threshold behaviour~\cite{crow2016improved}.
Related redundant-syndrome constructions were later developed in Refs.~\cite{premakumar20212,guttentag2024robust}.
These methods primarily protect the classical syndrome record; they do not, by themselves, solve the hook-error problem in data qubits studied here.

\noindent Yoder and Kim designed syndrome-extraction circuits for triangle codes that remain fault-tolerant under circuit-level noise, at the cost of additional qubits~\cite{Yoder2017surfacecodetwist}.
Chao and Reichardt later addressed this issue by introducing a more resource-efficient scheme that requires only two additional qubits, an ancilla qubit, and a flag qubit~\cite{PhysRevLett.121.050502}.
This approach is applicable to all stabilizer codes~\cite{chamberland2017flag,chao2020flag}.
Subsequent work produced several variants of the flag-qubit method, optimizing circuit depth by dynamically and adaptively selecting stabilizer measurements and introducing tools to simulate and optimize the flag approach~\cite{debroy2020extended,bhatnagar2023low,WOS:001379568200001,WOS:001317800000001,pato2024optimization}.
Lao and Almudever extended this approach by introducing a geometrical interaction mapping~\cite{PhysRevA.101.032333}.
They combine fault-tolerant flagging with quantum circuit mapping to create an efficient flag-bridge implementation of FT quantum error correction on near-term devices.

\noindent A further reduction in ancillary-qubit overhead was studied by Muyuan Li \emph{et al.}
They presented a $[[7,1,3]]$ non-CSS stabilizer code for which a suitable ordering of the data-ancilla interactions can correct hook errors caused by a single fault in the ancilla qubit~\cite{Muyuna}.
They evaluated the code under anisotropic and depolarizing noise models.
Under the anisotropic model considered in Ref.~\cite{Muyuna}, one bare ancilla per stabilizer measurement was sufficient to satisfy their fault-tolerance condition.
Under the standard two-qubit depolarizing model, however, additional flag qubits were required to detect hook errors.
Related ordering-based bare-ancilla circuits were subsequently developed for Bacon-Shor and compass codes~\cite{PhysRevA.98.050301,huang2020fault}.
Maheshwari \emph{et al.} analyzed a $[[8,1,3]]$ non-CSS code and reported its pseudo-threshold behavior under depolarizing and anisotropic noise models using bare-ancilla syndrome extraction~\cite{maheshwari2024fault}.

\noindent The works cited above analyze selected codes or code families.
They do not provide a general procedure for searching for suitable stabilizer generators, which would make the code fault-tolerant in the event of a fault in the ancilla qubit.
To address this gap, the next section formulates the conditions for BAC and an exact generator-and-ordering search for a given stabilizer set.

\color{black}

\section{Bare-ancilla syndrome extraction conditions and code search}
\label{sec: Bare construction}

\noindent This section states the conditions under which a given \([[n,k,3]]\) stabilizer code and its syndrome-extraction circuit form a valid BAC in the presence of a specified fault model.
For a given stabilizer set, the conditions specify how to search for suitable generators for BAC.
For each selected generator, it also searches every allowed data-ancilla interaction order.
Throughout this work, a BAC is specified by the triplet $(\mathcal C,\Pi,\mathcal M)$ in which $\mathcal C$ is the underlying stabilizer code, defined as:
\begin{align}
    \mathcal C &= \{g_1,\ldots,g_{n-k}\}, \nonumber
    \notag
\end{align}
Further, $\Pi$ is the set of permutations of the stabilizer operators, defined as:
\begin{align}
    \Pi & =    \{\pi_1,\ldots,\pi_{n-k}\},
    \notag
\end{align}
where each permutation $\pi_a$ is the data-ancilla interaction order used in measuring the stabilizer generator $g_a$.
We also use $\mathcal M$ to denote a predefined single-fault model.
When $\mathcal M$ is fixed by context, we suppress it from the notation and refer to $(\mathcal C,\Pi)$ as a BAC under $\mathcal M$.
Thus, any BAC never refers to the code $\mathcal C$ alone.

\noindent Any permutation of the measured stabilizer does not change the codespace.
However, it does change the hook errors propagated through the two-qubit gates present in the stabilizer measurement circuit, owing to the errors on the ancilla.
We therefore search for valid generators and interaction orders such that every hook error can be attributed to a unique syndrome. 
However, if the syndrome is the same across a set of hook errors, the errors are stabilizer-equivalent.

\subsection{Exact bare-ancilla fault-tolerance condition}
\label{subsec:exact-bac-condition}
% \vspace{-0.15cm}
\noindent  First, we define the condition that a given stabilizer code must satisfy to be FT under the chosen fault model during syndrome measurement.
Let $P,Q\in\mathcal P_n$ with binary representations $(\mathbf p_X\mid\mathbf p_Z)$ and $(\mathbf q_X\mid\mathbf q_Z)$, respectively then we define symplectic product as
\begin{equation*}
    \omega(P,Q)
   =\mathbf p_X^T\mathbf q_Z
    +\mathbf p_Z^T\mathbf q_X
    \pmod 2.
\end{equation*}
% where the commutation relation is defined as:
% \[
%  P Q
%    =(-1)^{\omega(P,Q)} Q P.
% \]
Let $\mathcal C$ be a $[[n,k,3]]$ stabilizer code generated by the stabilizer group $\mathcal S$.
Let $\mathcal S$ be generated by
$\{g_1,\ldots,g_{n-k}\}$. 
The Normalizer of $\mathcal S$ is 
\[
\mathcal N(\mathcal S)
   :=\left\{
       P\in\mathcal P_n:
       \omega(P,g)=0
       \text{ for every }g\in\mathcal S
     \right\}.
\]
\noindent 
The syndrome of $E\in\cP_n$ for a given generating set is 
\begin{equation*}
s_{\cS}(E) = \bigl(\omega(E,g_1),\ldots,\omega(E,g_{n-k})\bigr) \in\Ftwo^{n-k}.
\end{equation*}
\textcolor{black}{
For a BAC $(\mathcal{C},\Pi, \mathcal{M})$, we now define the set of all hook errors resulting from any error in the ancilla, during the stabilizer measurement round as $\mathcal{F}(\Pi, \mathcal{M})$.
We can now partition $\mathcal{F}(\Pi, \mathcal{M})$ into two disjoint sets $(\mathcal A, \mathcal B)$.
We define $\mathcal{A} \subset \mathcal{F}$, where for any error $E \in \mathcal{A}$ there exists $F \in \mathcal{A}$ such that $E^{\dagger}F \in \mathcal{S}$.
On the other hand $\mathcal B = \mathcal{F} \setminus \mathcal A$. 
We introduce the notation $\operatorname{Valid}_{\cS}(.)$ to validate the nature of such subsets of $\mathcal{F}$.
For instance the two cases can be summarized into $\operatorname{Valid}_{\cS}(\mathcal{A}) = 0$ and $\operatorname{Valid}_{\cS}(\mathcal{B}) = 1$  if every element in $\mathcal B$ have unique syndrome.}
% \textcolor{red}{Re-write-confusing:The other partition is a subset $\mathcal{B} \subset \mathcal{F}$, such that for all the members in the subset, there exists no other candidate from the same subset satisfying the previous condition.}
The FT condition of any BAC can be defined as follows:
\begin{definition}
    A given BAC $(\mathcal C,\Pi,\mathcal M)$ is FT if for $\mathcal{A}, \mathcal{B} \subset \mathcal{F}$, there exist unique syndromes for all the hook errors from the subset $\mathcal{B} \subset \mathcal{F} \implies \operatorname{Valid}_{\cS}(\mathcal{B}) = 1$.
    \label{def:general-bac}
\end{definition}
\noindent Definition \ref{def:general-bac} implies that for the subset $\mathcal{A} \subset \mathcal{F} : \operatorname{Valid}_{\cS}(\mathcal{A}) = 0$ unique syndrome attribution is not necessary.
This conditioning of FT BAC follows straightforwardly from the fact that hook errors differ by any stabilizer that affects the codespace in the same way.
Therefore, all such errors can have degenerate syndromes.
However, when any two errors differ by an operator from $\cN(\cS)\setminus\cS$, their action is not equivalent over the codespace, and therefore they cannot be attributed to the same syndrome.

\subsection{Hook-error models and syndrome-space bounds}
\label{subsec:hook-error-models}

\noindent In the previous section, we argued that any BAC is FT if all non-trivial hook errors are attributed to unique syndromes.
We now show that the availability of unique syndromes and the error-correcting capability of the BAC depend on the stabilizers weights.
Consider the error set consisting of all single qubit errors on the data qubits:
\begin{equation*}
\mathcal E_1 := \{I_i,X_i,Y_i,Z_i:1\leq i\leq n\}.
\end{equation*}
All the errors from $\mathcal E_1$ is correctable by a stabilizer code $\mathcal{S}$ with distance $\geq 3$.
We define the set of unique syndromes that can be attributed to all the errors in $\mathcal{E}_1$ as:\begin{align}
    \Sigma_1: =\{s_{\mathcal S}(E):E\in\mathcal E_1\}.\nonumber 
\end{align}
Therefore, the total number of syndromes offered by $\mathcal{S}$ that can be associated with the non-trivial hook errors are:
\begin{equation}
N_{\mathrm{free}}(\mathcal C) :=2^{n-k}- |\Sigma_1| .    
\label{eq:free syndrome of code}
\end{equation}

Consider a stabilizer generator $g$ measured in the interaction order $\pi$.
The corresponding ordered-permutation of $g$ is as follows: 
\begin{equation}
g = P_{a_1}P_{a_2}\cdots P_{a_{w_g}},
\qquad
P_{a_t}\in\{X_{a_t},Y_{a_t},Z_{a_t}\},
\label{eq:ordered-stabilizer}
\end{equation}
where $w_g=\operatorname{wt}(g)$ is the weight of $g$.
A Pauli-$X$ error on the ancilla immediately before the $t^{th}$ controlled-Pauli gate (corresponding to $P_{a_t}$) generates a hook error $\rho_{g}(t) = \prod_{\ell=t}^{w_g}P_{a_\ell},$ on the data qubits.
Here, $t=1,\ldots,w_g$ essentially is the time order of the permutation $\pi$.
Therefore, we define the set of all possible hook errors generated from $\pi$ ordered stabilizer $g$, due to any fault model $\mathcal{M}$ as $\mathcal H_{\mathcal M}(g,\pi)$.
Further, as all the errors from $\mathcal{E}_1$ already have unique syndromes, the only task that remains is to assign the same for all the hook errors affecting more than one data qubit.
Also, from our discussion in the previous section, it can be understood that from this remaining set of errors, we only need to assign a unique syndrome for any of the stabilizer equivalent error candidates, and for the rest of the equivalent hook error candidates, this becomes a degenerate assignment.
For instance, if any two hook errors $E$ and $F$, affecting data qubits, are stabilizer equivalent, that is, $E\sim_{\cS}F$, it means $E^{\dagger}F\in\cS$. 
The minimum weight of the stabilizer coset of $E$ is defined as $\operatorname{wt}_{\cS}(E) := \min_{S\in\cS}\operatorname{wt}(ES)$.
We define the set of all possible hook errors generated due to stabilizer $g$, such that its minimum stabilizer coset weight is greater than $1$ as:
\begin{equation}
\mathcal U_{\mathcal M}(g,\pi) := \left\{ E\in \mathcal H_{\mathcal M}(g,\pi): \operatorname{wt}_{\cS}(E)>1 \right\}.
\label{eq:additional-hook-error-set}
\end{equation}
\\

\noindent Consider the example in, Fig.~\ref{fig: measure syndrome faliure1}.
We show the propagated hook errors on the data qubits generated by single ancilla faults occurring at different locations during the ordered measurement of $g=P_3P_4P_1P_2P_5$.

\begin{figure}[ht]
    \centering
    \resizebox{0.485\textwidth}{!}{
    \begin{quantikz}[row sep=0.30cm]
\lstick[5]{data}&\lstick{$q_5$}\setwiretype{n}&\setwiretype{q}\slice[style={black!50, dashed}]{$a_1$}& \slice[style={black!50, dashed}]{$a_2$}&\slice[style={black!50, dashed}]{$a_3$}&\slice[style={black!50, dashed}]{$a_4$}&\slice[style={black!50, dashed}]{$a_5$}&\gate{\bm{P}}&&\lstick{$\textcolor{blue}{\bm{P_5}}$}\setwiretype{n}&\lstick{$\textcolor{teal!70}{\bm{P_5}}$}\setwiretype{n}&\lstick{$\textcolor{red!70}{\bm{P_5}}$}\setwiretype{n}&\lstick{$\textcolor{OliveGreen!70}{\bm{P_5}}$}\setwiretype{n}&\lstick{$\textcolor{violet!70}{\bm{P_5}}$}\setwiretype{n}\\
&\setwiretype{n}\lstick{$q_4$}\setwiretype{n}&\setwiretype{q}& &\gate{\bm{P}}&&&&&\lstick{$\textcolor{blue}{\bm{P_4}}$}\setwiretype{n}&\lstick{$\textcolor{teal!70}{\bm{P_4}}$}\setwiretype{n}&\lstick{$\textcolor{red!70}{\bm{I_4}}$}\setwiretype{n}&\lstick{$\textcolor{OliveGreen!70}{\bm{I_4}}$}\setwiretype{n}&\lstick{$\textcolor{violet!70}{\bm{I_4}}$}\setwiretype{n}\\
&\setwiretype{n}\lstick{$q_3$}&\setwiretype{q} &\gate{\bm{P}}&&&&&&\lstick{$\textcolor{blue}{\bm{P_3}}$}\setwiretype{n}&\lstick{$\textcolor{teal!70}{\bm{I_3}}$}\setwiretype{n}&\lstick{$\textcolor{red!70}{\bm{I_3}}$}\setwiretype{n}&\lstick{$\textcolor{OliveGreen!70}{\bm{I_3}}$}\setwiretype{n}&\lstick{$\textcolor{violet!70}{\bm{I_3}}$}\setwiretype{n}\\
&\setwiretype{n}\lstick{$q_2$}&\setwiretype{q}&&&&\gate{\bm{P}}&&&\lstick{$\textcolor{blue}{\bm{P_2}}$}\setwiretype{n}&\lstick{$\textcolor{teal!70}{\bm{P_2}}$}\setwiretype{n}&\lstick{$\textcolor{red!70}{\bm{P_2}}$}\setwiretype{n}&\lstick{$\textcolor{OliveGreen!70}{\bm{P_2}}$}\setwiretype{n}&\lstick{$\textcolor{violet!70}{\bm{I_2}}$}\setwiretype{n}\\
&\setwiretype{n}\lstick{$q_1$}&\setwiretype{q}&&& \gate{\bm{P}}&&&&\lstick{$\textcolor{blue!70}{\bm{P_1}}$}\setwiretype{n}&\lstick{$\textcolor{teal!70}{\bm{P_1}}$}\setwiretype{n}&\lstick{$\textcolor{red!70}{\bm{P_1}}$}\setwiretype{n}&\lstick{$\textcolor{OliveGreen!70}{\bm{I_1}}$}\setwiretype{n}&\lstick{$\textcolor{violet!70}{\bm{I_1}}$}\setwiretype{n}\\
&\setwiretype{n}\lstick{$\ket{+}$}& \setwiretype{q}\qwbundle{$\mathbf{X}$}&
\ctrl[style={{draw=blue!70, line width = 0.7mm}},label style=blue]{-3}&\ctrl[style={{draw=teal!70, line width = 0.7mm}},label style=teal]{-4}&\ctrl[style={{draw=red!70, line width = 0.7mm}},label style=red]{-1}&\ctrl[style={{draw=OliveGreen!70, line width = 0.7mm}},label style=OliveGreen]{-2}&\ctrl[style={{draw=violet!70, line width = 0.7mm}},label style=violet]{-5}&\meter{}&\setwiretype{n}\lstick{\textbf{X}}
\end{quantikz}
}
\caption{Hook-error propagation from an ancilla qubit due to $X$ fault during measurement of the ordered stabilizer $g=P_3P_4P_1P_2P_5$, where $P_i\in\{X_i,Y_i,Z_i\}$.
A fault at location $a_t$ propagates through the remaining controlled-Pauli gates and gives $\rho_g(t)$.
Matching colors identifies the fault location and its hook error.}
    \label{fig: measure syndrome faliure1}
\end{figure}

\noindent Now we discuss the two cases of hook errors for circuit-level
uncorrelated and correlated hook errors, and show how the correctability of these errors for BACs depends on
the weight $w_g$ of the stabilizer $g$.
\subsubsection{\textcolor{black}{Un-correlated errors: Affecting only ancilla qubit}}

\noindent 
The first error class we consider is a very simplified model
in which errors can occur only in the ancilla qubit at
any of the error locations described in the previous section.
We denote this fault model using $\mathcal M_{\mathrm a}$.
We previously discussed that there should be a set
of unassigned syndromes that can be further associated
with hook errors. This is essential to achieve fault tolerance against hook errors. We now show that the number of such abundant syndromes depends on the weight of the underlying stabilizers.
Consider the following lemma:
\begin{lemma}
\label{lemma:one qubit error}
For any stabilizer generator $g \in \mathcal{S} : g= \prod_{i} P_{a_i}$ of the $[[n,k,3]]$ stabilizer QECCs, if the weight of the stabilizer satisfies $w_g > 3$, then under the fault model $\mathcal M_{\mathrm a}$ the quantity $\mathcal U_{\mathcal M_{\mathrm a}}(g,\pi)$ satisfies:
\begin{equation*}
\left|\mathcal U_{\mathcal M_{\mathrm a}}(g,\pi)\right| \leq w_g-3.
\label{eq:ancilla-local-weight-bound}
\end{equation*}
\end{lemma}

\begin{proof}
Using the ordered form in Eq.~\eqref{eq:ordered-stabilizer}, a fault before the first interaction gives $\rho_g(1)=g\in\cS$.
A fault before the second interaction produces
\[
\rho_g(2)=P_{a_2}\cdots P_{a_{w_g}}=gP_{a_1}
\sim_{\cS}P_{a_1},
\]
whereas a fault before the final interaction leaves $\rho_g(w_g)=P_{a_{w_g}}$ on the data qubits.
These three hook errors are stabilizer-equivalent to Pauli errors of $\mathcal E_1$.
Therefore, only the $w_g-3$ locations can contribute elements to $\mathcal U_{\mathcal M_{\mathrm a}}(g,\pi)$.
Some of those propagated errors may also be stabilizer equivalent to errors in $\mathcal E_1$ or may coincide, so this is an upper bound.
\end{proof}
\begin{corollary}
\label{cor:ancilla-only-count}
For a given $[[n,k,3]]$ code, with stabilizer generator set \textcolor{black}{ $\mathcal{S} = \{g_1,\cdots,g_{n-k}\}$ with their corresponding interaction order $\{\pi_1,\cdots,\pi_{n-k}\}$}, the total number of hook errors belonging to distinct stabilizer cosets is given by
\begin{equation}
N_{\mathrm a} \leq \min\left\{ N_{\mathrm{free}}(\mathcal C), \sum_{g\in\mathcal S}\max\{w_g-3,0\} \right\}.
\label{eq:ancilla-only-capacity}
\end{equation}
if $\operatorname{Valid}_{\cS}
\bigl(\mathcal E_1\cup\mathcal U_{\mathrm a}\bigr)=1$,
where,
$\mathcal U_{\mathrm a}
:= \bigcup_{g\in\mathcal S} \mathcal U_{\mathcal M_{\mathrm a}}(g,\pi)$.

% \end{equation}
\end{corollary}
\begin{proof}
 The complete hook error set generated by $\mathcal M_a$ is $\mathcal E_1 \cup \mathcal U_{\mathrm a}$, so if we assume $\operatorname{Valid}_{\cS}
\bigl(\mathcal E_1\cup\mathcal U_{\mathrm a}\bigr)=1$, then from Definition~\ref{def:general-bac}, we can say for the fault model $\mathcal{M}_a$ all the errors in set $\mathcal E_1 \cup \mathcal U_{\mathrm a}$ must have unique syndrome for the BAC to be FT.
% Hence, all the distinct Pauli errors in $\mathcal U_{\mathrm a}$ have unique syndromes.
Therefore, using Eq.\eqref{eq:free syndrome of code}, we get
\[
N_{\mathrm a}\leq N_{\mathrm{free}}(\mathcal C).
\]
Further from Lemma~\ref{lemma:one qubit error}, we also get:
\[
N_{\mathrm a}
\leq
|\mathcal U_{\mathrm a}|
\leq
\sum_{g\in\mathcal S}\max\{w_g-3,0\},
\]
Together both these equations proves the corollary and Eq.~\eqref{eq:ancilla-only-capacity}.
\end{proof}

\subsubsection{\textcolor{black}{Correlated two-qubit gate errors: Affecting data and ancilla both due to faulty two-qubit gates}}
\label{sec:correlated error two qubit gate}

\noindent \textcolor{black}{We now consider a more practical error model.
We assume that the faulty two-qubit gates used for the stabilizer measurements introduce errors on both the data and ancilla qubits.}
\textcolor{black}{We denote this fault model using $\mathcal M_c$.}
\textcolor{black}{These faults can introduce arbitrary single-qubit Pauli errors on the data qubits, in addition to Pauli errors in the ancilla qubit.}
\textcolor{black}{Therefore, these faults generate a larger set of hook errors compared to the uncorrelated error model discussed previously.}
\textcolor{black}{We now establish a bound on the total number of unique syndromes required to achieve fault tolerance for a given $[[n,k,3]]$ code.}

% Let \(\mathcal M_{\mathrm c}\) denote the correlated ancilla-data Pauli fault model.
% A two-qubit Pauli error is applied immediately after the corresponding ideal controlled-Pauli interaction due to a fault in the ancilla.
% An \(I\) or \(Z\) Pauli error on the ancilla does not propagate through later interactions and therefore gives at most a single-qubit data error.
% An \(X\) or \(Y\) Pauli error can propagate through the remaining interactions.
\noindent \textcolor{black}{Every hook error in this model can be seen as 
\begin{align}
E=P_{a_1}\cdots P_{a_{t-1}}R_{a_t}, & \qquad 
t=2,\ldots,w_g-1, \nonumber \\ 
\quad & \text{where }R\in\{X,Y,Z\},
\label{eq:correlated-hook-representatives}
\end{align}
}
If a fault occurs in the starting or in the end, the corresponding propagated error lies in $\mathcal E_1$.
The cases with \(R=P_{a_t}\) is equivalent to the $\mathcal M_a$ fault model.
Therefore, the $\mathcal M_c$ model also includes the $\mathcal M_a$ fault model.

\begin{lemma}
\label{lemma:two qubit error}
For any stabilizer generator $g \in \mathcal{S}: g= \prod_{i} P_{a_i}$ of the \texorpdfstring{$[[n,1,3]]$}{[[n,k,3]]} stabilizer QECCs, if the weight of the stabilizer satisfies $w_g > 3$, then for the number of uncorrectable hook errors generated by the faults model $\mathcal M_c$,  the quantity $\mathcal U_{\mathcal M_{\mathrm c}}(g,\pi)$ satisfies:
% Let $g\in\cS$ have weight $w_g\geq3$ with interaction order $\pi$.
% Under the correlated ancilla-data fault model described above,
\begin{equation*}
\left|\mathcal U_{\mathcal M_{\mathrm c}}(g,\pi)\right|
\leq
3(w_g-2).
\end{equation*}
\end{lemma}

\begin{proof}
For a weight $w_g$, there are $w_g-2$ interior positions in Eq.~\eqref{eq:correlated-hook-representatives}, and at each position there are at most three nontrivial choices of $R$.
Removing those propagated errors which belong to the set $\mathcal E_1$ and duplicate Pauli operators can also be possible, which only reduces the size of $\mathcal U_{\mathcal M_{\mathrm c}}(g,\pi)$.
Therefore, its cardinality is at most $3(w_g-2)$.
\end{proof}

\begin{corollary}
\label{cor:correlated-count}
For a given $[[n,k,3]]$ code, with stabilizer generator set $\mathcal{S} = \{g_1,\cdots,g_{n-k}\}$ with their corresponding interaction order $\{\pi_1,\cdots,\pi_{n-k}\}$, the total number of hook errors belonging to distinct stabilizer cosets is given by
\begin{equation}
N_{\mathrm c} \leq \min\left\{ N_{\mathrm{free}}(\mathcal C), \sum_{g\in\mathcal S}3\max\{w_g-2,0\} \right\}.
\label{eq:correlated-capacity}
\end{equation}
if $\operatorname{Valid}_{\cS}
\bigl(\mathcal E_1\cup\mathcal U_{\mathrm c}\bigr)=1$,
where,
$\mathcal U_{\mathrm c}
:= \bigcup_{g\in\mathcal S} \mathcal U_{\mathcal M_{\mathrm c}}(g,\pi)$.

\end{corollary}

\begin{proof}
As in Corollary~\ref{cor:ancilla-only-count}, hook errors are omitted from $\mathcal U_{\mathrm c}$ whose stabilizer equivalent belongs to $\mathcal E_1$. So $\operatorname{Valid}_{\cS}
\bigl(\mathcal E_1\cup\mathcal U_{\mathrm c}\bigr)=1$  is therefore the exact condition in Definition~\ref{def:general-bac} for $\mathcal M_c$.
If it holds, the $N_{\mathrm c}$ distinct Pauli errors occupy distinct syndromes outside $\Sigma_1$, so $N_{\mathrm c}\leq N_{\mathrm{free}}(\mathcal C).$
The union bound and Lemma~\ref{lemma:two qubit error} give
\[
N_{\mathrm c} \leq |\mathcal U_{\mathrm c}| \leq \sum_{g\in\mathcal S}3\max\{w_g-2,0\},
\]
which proves Eq.~\eqref{eq:correlated-capacity}.
\end{proof}
\noindent 
Satisfying either  bounds is not sufficient for fault tolerance as $\operatorname{Valid}_{\cS}
\bigl(P,Q\bigr)=1$ needs to be satisfied, where $P,Q$ are the hook errors due to fault model $\mathcal{M}$.
In the next section, we present the algorithmic procedure for finding generators that satisfy the FT condition.

\subsection{Algorithm to find the $[[n, k, 3]]$ BACs}
\label{subsec:exact-basis-ordering-search}

\noindent We now describe a procedure for finding interaction orders that remain FT when all stabilizer generators are considered together.
For a given $[[n,k,3]]$ stabilizer code, the procedure either uses a fixed set of stabilizer generators or chooses a generating set from a larger pool of stabilizer set.
The syndrome-space bounds derived above provide a useful check, but the candidate is accepted only when the combined set of single-qubit errors and fault-induced hook errors satisfies Definition \ref{def:general-bac}.
Hence, we now describe an algorithm we designed to find a suitable permutation for each generator and for the whole stabilizer code that satisfies the FT conditions for BAC or Definition \ref{def:general-bac}.

\subsubsection{Local interaction-order search}
\noindent For $g\in\cS\setminus\{I\}$, let $\operatorname{Ord}(g)$ denote the allowed interaction orders of $g$.
For $\pi\in\operatorname{Ord}(g)$ and fault model $\mathcal M$, the set $\mathcal H_{\mathcal M}(g,\pi)$ 
is the corresponding hook error set.
And it must correct all the Pauli errors present in its set. 
Algorithm~\ref{alg:local-order-search} checks every allowed interaction order for $g$.
For each order $\pi$, it enumerates the complete set of hook errors for the chosen fault model and examines it jointly with $\mathcal E_1$.
The verification includes collisions of syndromes between two hook errors and collisions between a hook error and an error $E \in \mathcal E_1$.
A repeated nonzero syndrome is accepted when the two errors are stabilizer-equivalent; otherwise, it is rejected.

\noindent The output $\mathtt{Ord\_list_g}$ is the list of locally valid permutation of stabilizer generator $g$.
% Generators of weight two have no permutation search under the fault models considered here, but their propagated data-error sets are still included in the joint test.
For a weight-$w_g$ operator, the algorithm examines at most $w_g!$ interaction orders.
If the number of single-faults in the generators grows linearly with their corresponding weight $w_g$, the algorithm enumerates $\mathcal O(w_g!w_g)$ hook errors.
\vspace{-0.25cm}
{\linespread{1}\selectfont
\begin{algorithm}[ht]
\color{black}
\DontPrintSemicolon
\caption{Local interaction-order search}
\label{alg:local-order-search}

\KwIn{Stabilizer group $\cS$; stabilizer
$g\in\cS\setminus\{I\}$; fault model $\mathcal M$}

\KwOut{Candidate list $\mathtt{Ord\_list_g}$}

\BlankLine
$\mathtt{Ord\_list_g}\gets\varnothing$\;

\ForEach{$\pi\in\operatorname{Ord}(g)$}{
    $\mathcal H\gets\mathcal H_{\mathcal M}(g,\pi)$\;

    \If{$\operatorname{Valid}_{\cS}
    \bigl(\mathcal E_1\cup\mathcal H\bigr)=1$}{
        $\mathtt{Ord\_list_g}\gets
        \mathtt{ Ord\_list_g}\cup\{(g,\pi,\mathcal H)\}$\;
    }
}

\Return{$\mathtt{Ord\_list_g}$}\;
\end{algorithm}

\subsubsection{Global generators-and-ordering search}
\label{subsec:global-basis-ordering-search}

\noindent We now describe the procedure that considers the whole generator set and its ordering to find the BAC, summarized in Algorithm~\ref{alg:global-basis-ordering-search}.
Let $\mathcal G_{\mathrm{cand}} = \{g^{(1)},\ldots,g^{(N_{\mathcal G})}\} \subseteq\mathcal S\setminus\{I\}$ be set of stabilizer elements where $ N_{\mathcal G}=|\mathcal G_{\mathrm{cand}}|$.
With initial fix generators, $\mathcal G_{\mathrm{cand}}= \{g_1,\cdots,g_{n-k}\}$ and $N_{\mathcal G}=n-k$, whereas for an exhaustive generators search, $\mathcal G_{\mathrm{cand}}=\mathcal S\setminus\{I\}$ and $N_{\mathcal G}=2^{n-k}-1$.
The search constructs a generating set by selecting independent stabilizer elements from $\mathcal G_{\mathrm{cand}}$ together with their locally valid interaction orders.
It also maintains the full Pauli-error set generated by the selected stabilizer generators.
Each time a candidate is added, the search applies $\operatorname{Valid}_{\mathcal S}$ to the enlarged error set.
% It therefore allows only those hook errors with the distinct syndromes. 
\textcolor{black}{Once $\mathcal S_f$ contains $n-k$ independent elements that satisfy Definition~\ref{def:general-bac}, it returns the required generators with their ordering.}
Independence of the stabilizers is checked by $\operatorname{rank}$.
The search is initialized with $\mathcal F=\mathcal E_1$.
Every subsequent validity test, therefore, includes collisions with $\mathcal E_1$, collisions among propagated-data errors from the same stabilizer, and collisions between propagated errors from different stabilizers.
The global searches must be performed separately for each specified fault model $\mathcal M$.
{\linespread{1}\selectfont
\begin{algorithm}[ht]
\color{black}
\DontPrintSemicolon
\caption{Global generators-and-ordering search}
\label{alg:global-basis-ordering-search}

\KwIn{Stabilizer group $\cS$ of a specified $[[n,k,3]]$ code;
ordered candidate pool $\mathcal G_{\mathrm{cand}} =\{g^{(1)},\ldots,g^{(N_{\mathcal G})}\}$;
candidate lists
$\{\mathtt{Ord\_list_g}:g\in\mathcal G_{\mathrm{cand}}\}$
returned by Algorithm~\ref{alg:local-order-search}}

\KwOut{A witness $(\mathcal S_f,\Pi,\mathcal F)$, or $\varnothing$ if no witness exists}

\BlankLine

\SetKwFunction{Search}{Search}
\SetKwProg{Fn}{Procedure}{:}{}

\Fn{\Search{$q,\mathcal S_f,\Pi,\mathcal F$}}{

    \If{$|\mathcal S_f|= n-k $}{
        \Return{$(\mathcal S_f,\Pi,\mathcal F)$}\;
    }

    \For{$\ell\gets q$ \KwTo $N_{\mathcal G}$}{
    $g\gets g^{(\ell)}$\;

        \If{$
        \operatorname{rank}(\mathcal S_f\cup\{g\})
        \neq |\mathcal S_f|+1$}{
            \textbf{continue}\;
        }

        \ForEach{$(g,\pi,\mathcal H)\in\mathtt{Ord\_list_g}$}{
            $\mathcal F'\gets\mathcal F\cup\mathcal H$\;

            \If{$\operatorname{Valid}_{\cS}(\mathcal F')=0$}{
                \textbf{continue}\;
            }

            $\mathcal S'_f\gets\mathcal S_f\cup\{g\}$\;
            $\Pi'\gets\Pi\cup\{(g,\pi)\}$\;

            $\mathsf{out}\gets$\Search{$\ell+1,\mathcal S'_f,\Pi',\mathcal F'$}\;

            \If{$\mathsf{out}\neq\varnothing$}{
            \Return{$\mathsf{out}$}\;
                }
            }
        }

    \Return{$\varnothing$}\;
}

\BlankLine
\Return{\Search{$1,\varnothing,\varnothing,\mathcal E_1$}}\;

\end{algorithm}
}

\noindent Suppose that every candidate stabilizer has weight at most $w_{\max}$.
Final set should contain $n-k$ independent stabilizers from $|\mathcal N_\mathcal G|$ stabilizers, and each selected stabilizer has at most $w_{\max}!$ interaction orders.
A worst-case upper bound to find the valid global generators and ordering is 
\begin{equation}
\binom{N_{\mathcal G}}{n-k}
\bigl(w_{\max}!\bigr)^{n-k}.
\label{eq:global-search-branch-bound}
\end{equation}
Independence tests, local rejection, and collision pruning can only reduce this number.
\noindent
\textcolor{black}{The next section, i.e., Sec .~\ref {sec:general-nk3-extensions}, gives a construction that extends a distance-three BAC with a verified BAC to larger block lengths and, through blockwise composition, to larger numbers of encoded qubits.}

\section{General extensions of distance three bare ancilla codes}
\label{sec:general-nk3-extensions}

\noindent
This section establishes two closure operations for BAC. 
Throughout this section, the single ancilla fault model $\mathcal M$ is fixed.
The shorthand $[[n,k,3]]_{\mathrm{BAC}}$ denotes a pair $(\mathcal C,\Pi)$ under $\mathcal M$, where $\mathcal C$ has parameters $[[n,k,3]]$ with corresponding error set $\mathcal F(\Pi, \mathcal M)$.
\textcolor{black}{The first construction adds  $r$ physical qubits, while preserving logical information and minimum distance,}
\begin{equation*}
[[n,k,3]]_{\mathrm{BAC}}
\longrightarrow
[[n+r,k,3]]_{\mathrm{BAC}}.
\label{eq:fixed-k-extension-map}
\end{equation*}
% \textcolor{red}{The above map is not defined properly, what is $r$?}
The second construction places two BAC code blocks on disjoint data qubits, thereby increasing the logical information. Combined with the first construction, it gives a code for every sufficiently large physical length at any prescribed $k'>k$, provided that a valid $[[n,1,3]]$  BAC is available under the same fault model.
Both constructions are for a general stabilizer code that satisfies the BAC property.

\subsection{Pendant extension: $[[n,k,3]] \longrightarrow \ [[n+r,k,3]]$ }
\label{subsec:pendant-extension}

\noindent
\textcolor{black}{Let $\mathcal C$ be an $[[n,k,3]]$ stabilizer code, also act as parent code  with $k\geq 1$ and stabilizer group $\mathcal S$.}
By the definition of distance, this code has a logical Pauli operator
\begin{equation}
L_0\in\mathcal N(\mathcal S)\setminus\mathcal S \text{ such that } 
\operatorname{wt}(L_0)=3.
\label{eq:weight-three-logical}
\end{equation}
Now consider a physical qubit $h\notin\operatorname{supp}(L_0)$, where $``\operatorname{supp}"$ denotes the support of operator $L_0$, implying $(L_0)_h = I_h$.
Such a qubit exists because the quantum Singleton bound ~\cite{CerfCleve1997}.
This bound gives \(n-k\geq 2(d-1)=4\).
Since \(k\geq1\), we have \(n\geq5\), whereas \(|\operatorname{supp}(L_0)|=3\).

\noindent \textcolor{black}{Let $J_r$ is the set which contains $r$ new physical qubits.} 
% \textcolor{red}{re-word:Let $J_r$ be the set of new $r$ additional encoding data qubit set, disjoint from the original data qubit set.}
We also define unitary $U_r$ and the new codespace $\mathcal C^{(r)}$ as:
\begin{equation}
U_r=\prod_{\ell\in J_r}\operatorname{CZ}_{h,\ell},
\qquad
\mathcal C^{(r)}
=
U_r\bigl(\mathcal C\otimes
\lvert+\rangle^{\otimes r}\bigr).
\label{eq:pendant-extension}
\end{equation}
\textcolor{black}{The extra qubits added to the parent code, looks like pendants being attached to one of the qubits from the parent code.}
Hence we call the extension $\mathcal{C} \rightarrow \mathcal{C}^{(r)}$ as the \emph{pendant} extension model.
\begin{proposition}
\label{prop:pendant-stabilizers}
\textcolor{black}{For any parent $[[n,k,3]]$ code $\mathcal C$ with stabilizer generators $\{ g_1,\ldots,g_{n-k}\}$}, the stabilizer generator set of the extended code family $\mathcal C^{(r)}$ due to the encoding operation by $U_r$ with reference to a single physical qubit $h\notin\operatorname{supp}(L_0)$ is given by
\begin{align}
g_a^{(r)} &= g_a\prod_{\ell\in J_r}Z_\ell^{\gamma_a},
&&a=1,\ldots,n-k,
\label{eq:extended-old-generators}
\\
g_\ell &= Z_hX_\ell,
&&\ell\in J_r.
\label{eq:new-pendant-generators}
\end{align}
\textcolor{black}{where $\gamma_a=\omega(g_a,Z_h)$.}
% \textcolor{red}{What is $\gamma_a$?}

\noindent If $L\in\mathcal N(\mathcal S) $ is a logical operator of $\mathcal{C}$
% \textcolor{red}{confusing!with none of its support belonging to $h\notin\operatorname{supp}(L_0)$, }
% \textcolor{green}{This is any logical from $\mathcal N(\mathcal S)$ , not $L_0$(which has the property that $wt(L_0) = 3$) and we show how that logical operator evolves when a unitary is applied. In this proposition, we just define the stabilizer and logicals of the extended code. }
then those logical operators transform to the following logical for the code $\mathcal C^{(r)}$:
\begin{equation}
L^{(r)} = U_rLU_r^\dagger = L\prod_{\ell\in J_r} Z_\ell^{\omega(L,Z_h)}.
\label{eq:extended-logical-operator}
\end{equation}
\end{proposition}
\begin{proof}
% We first reiterate the notation for checking the commutativity of two operators is $\gamma_a=\omega(g_a,Z_h)$.
% \textcolor{red}{$\gamma_a$ is coming here! WOW!}
The new qubits $\ell \in J_r$ initialized with $\ket{+}$ state.
Therefore, the extended stabilizer group is
\begin{equation*}
\widetilde{\mathcal S}^{(r)} = \left\langle \mathcal S, X_\ell:\ell\in J_r \right\rangle.
\end{equation*}
Now the proposed stabilizer generators in Eq.~\eqref{eq:extended-old-generators} and
Eq.~\eqref{eq:new-pendant-generators} follow from the following rules:
\begin{align}
\operatorname{CZ}_{h,\ell}X_h
\operatorname{CZ}_{h,\ell}&=X_hZ_\ell, \nonumber
\\ \nonumber
\operatorname{CZ}_{h,\ell}Z_h
\operatorname{CZ}_{h,\ell}&=Z_h.\\ \nonumber
\operatorname{CZ}_{h,\ell}X_\ell
\operatorname{CZ}_{h,\ell}&=Z_hX_\ell, \nonumber
\end{align}
\textcolor{black}{The first two conditions give Eq.~\eqref{eq:extended-old-generators}}
The third rule gives Eq. \eqref{eq:new-pendant-generators}.
\textcolor{black}{Therefore, the stabilizers of the parent code contribute $n-k$ independent generators, and the codespace extension to the new $r$ qubits contributes $r$ independent generators}.
The total number of stabilizer generators of $\mathcal C^{(r)}$ is $\widetilde{\mathcal S}^{(r)}$ is $(n+r)-k$.
Under these conjugation rules the new logical operators also assume the expression of Eq.~\eqref{eq:extended-logical-operator}.
\end{proof}
\begin{theorem}
\label{thm:pendant-parameters}
For a single $h\notin\operatorname{supp}(L_0)$ the new stabilizer code $\mathcal C^{(r)}$ derived from $\mathcal{C}$, with the additional $r$ physical qubits using the unitary $U_r$, is a $[[n+r,k,3]]$ code, for every $r\geq0$.
\end{theorem}
\begin{proof}
Proposition~\ref{prop:pendant-stabilizers} shows that $\mathcal C^{(r)}$ encodes exactly $k$ logical qubits.
Since $(L_0)_h=I_h$, Eq.~\eqref{eq:extended-logical-operator} gives $L^{(r)}=L_0$.
Therefore the weight $3$ logical operator of $\mathcal{C}$, i.e., $L_0$ remains a logical operator of $\mathcal C^{(r)}$.
This implies that
\begin{equation}
d\bigl(\mathcal C^{(r)}\bigr)\leq 3.
\label{eq:pendant-distance-upper}
\end{equation}
Now consider, $L^{(r)}$ and apply the following unitary transformation:
$\widetilde L^{(r)}=U_r^\dagger L^{(r)} U_r$.
These operators correspond to the logical of the intermediate stabilizer set $\widetilde{\mathcal S}^{(r)}$, i.e., $\widetilde L^{(r)} \in \mathcal N\bigl(\widetilde{\mathcal S}^{(r)}\bigr) \setminus \widetilde{\mathcal S}^{(r)}.$
Since \(\omega(\widetilde L^{(r)},X_\ell)=0\) for every \(\ell\in J_r\), we can state that the operator of \(\widetilde L^{(r)}\) on each new qubit from $J_r$ is either \(I_\ell\) or \(X_\ell\).
Hence, the intermediate logical operators, before the encoding unitary of the extended code, satisfy:
\begin{equation}
\widetilde L^{(r)} = L\prod_{\ell\in J_X}X_\ell,
\label{eq:pulled-back-logical}
\end{equation}
where, $J_X\subseteq J_r$.
% Now, if we define the two categories for the support operator type of the logical operators $L$, using the following:
% \begin{equation*}
% \chi_{h}(L)=
% \begin{cases}
% 0,&L_h\in [I_h,Z_h],\\
% 1,&L_h\in [X_h,Y_h].
% \end{cases}
% \end{equation*}
The unitary encoding of $\mathcal{C}^({r})$ along with the operator evolution rule-set used in the previous proposition gives the new logical operators ($\widetilde L^{(r)} \rightarrow L^{(r)}$) as:
\begin{equation*}
L^{(r)} = L\ Z_h^{|J_X|} \left( \prod_{\ell\in J_r}Z_\ell^{\omega(L,Z_h)} \right) \left( \prod_{\ell\in J_X}X_\ell \right).
\label{eq:extended-logical-weight-form}
\end{equation*}
If $L_h=I_h$,  then $h\notin\operatorname{supp}(L)$.
This implies that the factor $Z_h^{|J_X|}$ is either $I_h$ or $Z_h$.
\textcolor{black}{Therefore, along with operators in $L$, either we will get an operator on $l \in J_X$ qubits or not. 
In both cases, it cannot remove the other supports already present in $L$.}
% \textcolor{red}{Confusing:Therefore, it can either preserve or remove the non-trivial logical operator type on the $h^{\mathrm{th}}$ qubit}.
If $L_h=Z_h$ and $|J_X|$ is even, this factor acts trivially at $h$.
% , so all old-qubit support is preserved.
If $L_h=Z_h$ and $|J_X|$ is odd, the factor cancels the $h$ component of $L$ and removes the single qubit support on $h$.
However, $|J_X|\geq1$, and the factors $X_\ell$ introduce $|J_X|$ new-qubit supports.
Hence, the weight cannot decrease.
If $L_h\in\{X_h,Y_h\}$, multiplication by $Z_h$ does not remove the support at $h$, and each new qubit carries either $Z_\ell$ or $Y_\ell$.
Thus, in every case,
\begin{equation*}
\operatorname{wt}(L^{(r)})\geq\operatorname{wt}(L)\geq 3.
\end{equation*}
Therefore $d(\mathcal C^{(r)})\geq 3$, which together with Eq.~\eqref{eq:pendant-distance-upper} proves the theorem.
\end{proof}

\subsection{Preservation of the bare-ancilla condition}
\label{subsec:pendant-bac-preservation}

\noindent 
We now discuss how the pendant extended $[[n+r,k,3]]$ code preserves the BAC nature similar to its parent code.
We consider a simpler scenario of the pendant extension, where $r=1$, i.e., $J_1=\{\ell\}$ and $\mathcal C^{(1)} = \operatorname{CZ}_{h,\ell} \bigl(\mathcal C\otimes\lvert+\rangle_\ell\bigr)$.
Let $\mathcal S^{(1)}$ denote the stabilizer group of $\mathcal C^{(1)}$.
By Proposition~\ref{prop:pendant-stabilizers}, its generators are
\begin{align}
g_a^{(1)} &= g_aZ_\ell^{\gamma_a}, \nonumber
&&a=1,\ldots,n-k,\\
g_\ell &= Z_hX_\ell.
\label{eq: generator of new code}
\end{align}

\noindent Let $\pi_a=(v_{1},\ldots,v_{w})$ be the data ancilla interaction order used to measure the parent stabilizer $g_a$, where $w=\mathrm{wt}(g_a)$.
For the extended stabilizer $g_a^{(1)}=g_aZ_\ell^{\gamma_a}$,
we choose the interaction order
\begin{equation}
\pi_a^{(1)}=
\begin{cases}
\pi_a, & \gamma_a=0,\\[2mm]
(v_{1},\ldots,v_{w},Z_\ell), & \gamma_a=1.
\end{cases}
\end{equation}
Thus, when $\gamma_a=1$, the original data-ancilla interaction order is preserved,
and the additional controlled-$Z_\ell$ interaction is applied last.

\noindent The new stabilizer $g_\ell$ is the two-weight generator.
Hence, the order of interaction does not matter.
We denote the resulting complete set of interaction orders by
\begin{equation}
\Pi^{(1)} = \{\pi_1^{(1)},\ldots, \pi_{n-k}^{(1)},\pi_\ell^{(1)}\}.
\label{eq: new code stabilizer ordering}
\end{equation}
where $\pi_l$ is the interaction order of $g_l$.

The proof for the preservation of the bare condition of $\mathcal{C}^{(1)}$ requires the following closure properties for the hook errors generated due to the new stabilizers under any fault model.
They state precisely what is needed from the circuit-level fault model.

\subsubsection*{Three closure properties}
\label{colsure_proeprties}

\begin{enumerate}
\item Any hook errors generated due to faults affecting the two qubit gates corresponding to the sub-part $g_a$ of the stabilizer operator $g^{(1)}_a$ has the form $EZ_\ell^\epsilon$, where $E\in\mathcal F(\Pi,\mathcal M)$ and $\epsilon\in\mathbb F_2$.
\item Any fault affecting the two qubit gates corresponding to the term due to the pendant extension of the stabilizers $g_a \rightarrow g^{(1)}_a$ generates hook errors supported only over the physical data qubits $\ell$.
\item Every propagated hook error generated while measuring $g_\ell$ is identity or, a single-qubit Pauli error up to any stabilizer-equivalence from the generating set $\mathcal S^{(1)}$.
\end{enumerate}

\begin{lemma}
\label{lem:pendant-circuit-closure}
For the uncorrelated fault model $\mathcal M_{\mathrm a}$ and the correlated fault model $\mathcal M_{\mathrm c}$, every single fault in the extended code $\mathcal{C}^{(1)}$ whose stabilizer generators are defined in Eq.\eqref{eq: generator of new code} satisfies all three closure properties listed above, provided the interactions follow the stated order $\Pi^{(1)}$ defined in Eq.\eqref{eq: new code stabilizer ordering}.
\end{lemma}

\begin{proof}
A fault in the extended code leaves the same error $E\in\mathcal F(\Pi,\mathcal M)$ on the original data qubits as in the parent circuit.
Thus, the final propagated data error is $EZ_\ell^\epsilon$, where $\epsilon \in (0,1)$.
A fault at the appended final gate gives the hook error, which is confined to $\ell$.
And $g_{\ell}$ is the two-weight stabilizer generator, which is always correctable, as it induces a single-qubit fault. 
Hence, the resulting ordering satisfies the closure properties. 
\end{proof}
\color{black}

% \begin{lemma}
% \label{lem:pendant-circuit-closure}
% For  the un-correlated and correlated fault models $\mathcal M_{\mathrm a}$ and $\mathcal M_{\mathrm c}$, the hook errors propagated from the stabilizer measurement circuits of the extended stabilizers due to the underlying circuit-level faults follows the \textit{three closure properties}.
% provided the interaction order of the stabilizers follow $\Pi^{(1)}$.
% \end{lemma}
% }
% \begin{proof}
% Proof of this lemma is fairly straightforward.
% A fault in the extended code leaves the same error $E$ on the original data qubits as in the seed circuit.
% Thus, the final hook error is $EZ_\ell^\epsilon$.
% A fault at the appended final gate gives the hook error, which is confined to $\ell$.
% And $g_{\ell}$ is the two-weight stabilizer generator, which is always correctable, as it induces a single-qubit fault. 
% \end{proof}
\begin{lemma}
\label{lem:lifted-error-validity}
Consider a FT BAC $(\mathcal{C}, \Pi, \mathcal{M})$ derived from the parent code $\mathcal{C}$ and all the possible hook errors generated due to faults from $\mathcal{M}$ is $\mathcal{F}(\Pi, \mathcal{M})$.
Therefore, the following hook error set $\mathcal{A}^{(1)}$ satisfy $\operatorname{Valid}_{\mathcal S^{(1)}}(\mathcal A^{(1)})=1$.
\begin{equation}
\mathcal A^{(1)} = \left\{ EZ_\ell^x: E\in\mathcal F, \ x \in\mathbb F_2 \right\} \cup \left\{ X_\ell^uZ_\ell^v: u,v\in\mathbb F_2 \right\}.
\label{eq:lifted-error-cover}
\end{equation}
\end{lemma}
\begin{proof}
Hook errors generated due to the stabilizers $g^{(1)}$ and $g_l$ from $\mathcal{S}^{(1)}$ fall into two categories, $ \left\{ EZ_\ell^x: E\in\mathcal F, \ x\in\mathbb F_2 \right\}$ and $\left\{ X_\ell^uZ_\ell^v: u,v\in\mathbb F_2 \right\}$.
Let $\boldsymbol \gamma=(\gamma_1,\ldots,\gamma_{n-k}) =s_{\mathcal S}(Z_h).$
With respect to the stabilizer generators $\{g_1^{(1)},\ldots,g_{n-k}^{(1)},g_\ell\}$, these two types of errors can generate the following syndromes:
\begin{align}
s_{\mathcal S^{(1)}}(EZ_\ell^x) &= \bigl(s_{\mathcal S}(E), \omega(E,Z_h)+x\bigr), \nonumber \\
s_{\mathcal S^{(1)}}(X_\ell^uZ_\ell^v) &= (u\boldsymbol\gamma,v). \nonumber
\end{align}
\noindent We consider three possible cases, where multiple errors can contribute to same syndrome.\\
First, consider that two propagated errors, $EZ_\ell^x$ and $FZ_\ell^y$, have the same extended syndrome.
From the closure properties, both the errors must have same bit values for the $n-k$ measured stabilizers for $\mathcal{S}$, i.e., $s_{\mathcal S}(E)=s_{\mathcal S}(F)$.
Since the stabilizer measurement circuit is derived from a FT BAC, such syndrome overlap between two errors imply: $D=E^\dagger F\in\mathcal S$.\\\\
% \textcolor{red}{What is $\alpha_a$?}
\textcolor{black}{Write $D=\prod_{a=1}^{n-k}g_a^{\alpha_a}$ where $\alpha_a \in \{0,1\}$.}
Equality of the final syndrome bits gives
\begin{equation*}
x+y = \omega(D,Z_h) = \sum_{a=1}^{n-k}\alpha_a\gamma_a \pmod 2.
\end{equation*}
Consequently,
\begin{equation*}
(EZ_\ell^x)^\dagger(FZ_\ell^y) = \prod_{a=1}^{n-k}(g_a^{(1)})^{\alpha_a} \in\mathcal S^{(1)}.
\label{eq:two-lifted-errors-safe}
\end{equation*}
Second, suppose a propagated error $EZ_\ell^x$ and an error $X_\ell^uZ_\ell^v$ have the same syndrome, then on comparing first $n-k$ bits we get
\begin{equation*}
s_{\mathcal S}(E) =u\boldsymbol\gamma =s_{\mathcal S}(Z_h^u).
\end{equation*}
Both $E$ and $Z_h^u$ belong to $\mathcal F$, because
$\mathcal E_1\subseteq\mathcal F$. Definition~\ref{def:general-bac}
therefore gives $D=E^\dagger Z_h^u\in\mathcal S.$
Writing $D=\prod_a g_a^{\alpha_a}$ and comparing the final syndrome bit gives
\begin{equation*}
x+v = \omega(D,Z_h) = \sum_a\alpha_a\gamma_a \pmod 2.
\end{equation*}
It follows that
\begin{equation*}
(EZ_\ell^x)^\dagger (X_\ell^uZ_\ell^v) = \left[\prod_a(g_a^{(1)})^{\alpha_a}\right]g_\ell^u \in\mathcal S^{(1)}.
\label{eq:mixed-errors-safe}
\end{equation*}
Third, suppose two errors supported on $\ell$ have the same syndrome.
Their product lies in $\mathcal N(\mathcal S^{(1)})$ and has weight at most one.
Theorem~\ref{thm:pendant-parameters} proves minimum distance of $\mathcal C^{(1)}=3$, so this product cannot be a nontrivial logical operator and must belong to $\mathcal S^{(1)}$.
All possible collisions in $\mathcal A^{(1)}$ are therefore safe.
\end{proof}

\begin{theorem}
\label{thm:pendant-bac-preservation}
Let $(\mathcal C,\Pi)$ represent the  $[[n,k,3]]_{\mathrm{BAC}}$ under $\mathcal M$, with $k\geq 1$.
Choose $L_0$ and $h$ such that $wt(L_0) = 3$ and $h \notin \operatorname{supp}(L_0)$.
If $\mathcal M$ satisfies the closure properties stated in Section~\ref{colsure_proeprties}, then $\bigl(\mathcal C^{(1)},\Pi^{(1)}\bigr)$ represent the $[[n+1,k,3]]_{\mathrm{BAC}}$ under the same fault model $\mathcal M$.
\end{theorem}

\begin{proof}
The parameters of $\mathcal C^{(1)}$ follow from Theorem~\ref{thm:pendant-parameters}.
By the closure properties, every error in $\mathcal F(\Pi^{(1)},\mathcal M)$ is stabilizer-equivalent to an element of $\mathcal A^{(1)}$.
Lemma~\ref{lem:lifted-error-validity} therefore gives
\[
\operatorname{Valid}_{\mathcal S^{(1)}}
\bigl(\mathcal F(\Pi^{(1)},\mathcal M)\bigr)=1.
\]
Hence $(\mathcal C^{(1)},\Pi^{(1)})$ satisfies Definition~\ref{def:general-bac} under $\mathcal M$.
\end{proof}

\begin{corollary}
\label{cor:repeated-pendant-extension}
Under the hypotheses of Theorem~\ref{thm:pendant-bac-preservation}, the construction be repeated at the same hub $h$.
For every $r\geq 0$, it gives an
\begin{equation*}
[[n+r,k,3]]_{\mathrm{BAC}}
\end{equation*}
code under $\mathcal M$.
\end{corollary}

\begin{proof}
The logical operator $L_0$ is unchanged by every extension, so it continues to have weight three and to act trivially on $h$ and on all new qubits.
Theorem~\ref{thm:pendant-parameters} and Theorem~\ref{thm:pendant-bac-preservation} therefore be applied inductively.
\end{proof}

\subsection{Blockwise composition for multiple logical qubits}
\label{subsec:blockwise-composition}

\noindent
The pendant extension preserves $k$.
To increase the number of encoded qubits, we combine two  BACs on disjoint sets of physical qubits.

\noindent Let $(\mathcal C_A,\Pi_A)$ and $(\mathcal C_B,\Pi_B)$ have stabilizer groups $\mathcal S_A$ and $\mathcal S_B$, respectively.
We define their blockwise composition by
\begin{equation}
(\mathcal C_A,\Pi_A)\odot(\mathcal C_B,\Pi_B)
:=
(\mathcal C_A\otimes\mathcal C_B,\Pi_A\sqcup\Pi_B).
\label{eq:blockwise-code-circuit-product}
\end{equation}
Here $\Pi_A\sqcup\Pi_B$ denotes the disjoint union of the two BAC interaction orders.
The stabilizer group of the composed code is
\begin{equation*}
\mathcal S_{AB}
:= \left\langle S_A\otimes I_B:S_A\in\mathcal S_A,\quad I_A\otimes S_B:S_B\in\mathcal S_B \right\rangle,
\label{eq:blockwise-stabilizer-product}
\end{equation*}
where $I_A$ and $I_B$ are the identity operators on blocks $A$ and $B$, respectively.
The symbol $\odot$ denotes the composition of the codes.
Every stabilizer generator is block-local: no controlled gate couples the two data blocks.

\begin{theorem}
\label{thm:blockwise-bac-composition}
Let $(\mathcal C_A,\Pi_A)$ and $(\mathcal C_B,\Pi_B)$ under the same single-fault model $\mathcal M$, represent $\mathcal C_A=[[n_A,k_A,3]]_{BAC},$ and $\mathcal C_B=[[n_B,k_B,3]]_{BAC}$ respectively.
Then their blockwise composition is $[[n_A+n_B,k_A+k_B,3]]_{\mathrm{BAC}}$ under $\mathcal M$.
\end{theorem}

\begin{proof}
Number of generators in $\mathcal S_{AB} = (n_A+n_B)-(k_A+k_B)$.
Every Pauli operator on the combined set of data qubits can be written as $P_A\otimes P_B$.
It lies in \(\mathcal N(\mathcal S_{AB})\) as \(P_A\in\mathcal N(\mathcal S_A)\) and \(P_B\in\mathcal N(\mathcal S_B)\).
Therefore,
\begin{equation*}
d(\mathcal C_A\otimes\mathcal C_B) = \min\{d(\mathcal C_A),d(\mathcal C_B)\} =3.
\end{equation*}
It remains to be verified whether the BAC condition holds.
Because all generators are block-local, a hook error arises because a single ancilla fault is confined to one block.
Let $E$ and $F$ be two errors with the same syndrome.
If both are supported in block $A$, the BAC condition for $(\mathcal C_A,\Pi_A)$ gives $E^\dagger F\in\mathcal S_{AB}$; the same argument holds for block $B$.

\noindent If the errors are in different blocks, write
\[
E=E_A\otimes I_B,
\qquad
F=I_A\otimes F_B.
\]
Equality of their syndromes implies
\[
s_{\mathcal S_A}(E_A)= 0,
\qquad
s_{\mathcal S_B}(F_B)= 0.
\]
Since the identity belongs to both local fault sets, applying the local
BAC conditions to $(I_A,E_A)$ and $(I_B,F_B)$ gives
\[
E_A\in\mathcal S_A,
\qquad
F_B\in\mathcal S_B.
\]
Consequently,
\[
E^\dagger F
=
E_A^\dagger\otimes F_B
\in\mathcal S_{AB}.
\]
This also includes the case in which one error is the identity.
Thus every equal-syndrome pair differs by an element of $\mathcal S_{AB}$, so the composed pair satisfies the BAC condition.
\end{proof}

\noindent  For a positive integer $q$, the notation $(\mathcal C,\Pi)^{\odot q}$ denotes the $q$-fold blockwise composition of $(\mathcal C,\Pi)$ with itself.

\begin{theorem}
\label{thm:prescribed-k-extension}
Let \((\mathcal C,\Pi)\) for \([[n,k,3]]_{\mathrm{BAC}}\) , and  \((\mathcal C_0,\Pi_0)\)  for
\([[n_0,1,3]]_{\mathrm{BAC}}\) under the same pendant-compatible single-fault model \(\mathcal M\).
For an integer \(k'>k\), define
\[
\nu:=k'-k,
\qquad
N_0(k'):=n+n_0\nu,
\]
and form
\[
(\mathcal C_{k'},\Pi_{k'})
 :=(\mathcal C,\Pi)\odot
   (\mathcal C_0,\Pi_0)^{\odot\nu}.
\]
Then \((\mathcal C_{k'},\Pi_{k'})\) represent \([[N_0(k'),k',3]]_{\mathrm{BAC}}\) under \(\mathcal M\).
Consequently, for every integer \(n'\geq N_0(k')\), there exists an \([[n',k',3]]_{\mathrm{BAC}}\)  under \(\mathcal M\).
\end{theorem}

\begin{proof}
Repeated application of Theorem~\ref{thm:blockwise-bac-composition} gives the blockwise pair with parameters \([[N_0(k'),k',3]]\).
For \(n'\geq N_0(k')\), set \(r:=n'-N_0(k')\).
Since code is distance three, then with weight three logical operator choose a hub outside its support and apply Corollary~\ref{cor:repeated-pendant-extension} exactly \(r\) times.
The resulting BAC is \([[N_0(k')+r,k',3]]=[[n',k',3]]\).
\end{proof}

\begin{corollary}
\label{cor:model-specific-seed-bounds}
Let $(\mathcal C,\Pi)$ pair represent $[[n,k,3]]_{\mathrm{BAC}}$.
\begin{enumerate}
\item Under the ancilla-only model $\mathcal M_{\mathrm a}$, the $[[6,1,3]]_{\mathrm{BAC}}$ in Appendix~\ref{app:six-qubit-certificate}  be used as the parent in Theorem~\ref{thm:prescribed-k-extension}.
Hence, for every $k'>k$, a BAC exists for every
\[
n'\geq n+6(k'-k),
\]
with parameters $[[n',k',3]]_{\mathrm{BAC}}$ under $\mathcal M_{\mathrm a}$.

\item Under the correlated ancilla-data model $\mathcal M_{\mathrm c}$, the $[[7,1,3]]_{\mathrm{BAC}}$ in Appendix~\ref{app:seven-qubit-certificate} be used as the parent.
Hence, for every $k'>k$, a BAC exists for every
\[
n'\geq n+7(k'-k),
\]
with parameters $[[n',k',3]]_{\mathrm{BAC}}$ under $\mathcal M_{\mathrm c}$.
\end{enumerate}
\end{corollary}

\noindent The choices $n_0=6$ under $\mathcal M_{\mathrm a}$ and $n_0=7$ under $\mathcal M_{\mathrm c}$ have minimum length within the enumerated one-message graph-code domain.
Nevertheless, the bound
\[
N_0(k')=n+n_0(k'-k)
\]
is a constructive sufficient bound and is not claimed to be the globally minimum length for an $[[n',k',3]]_{\mathrm{BAC}}$.

\noindent If the construction is required to contain at least one copy of $(\mathcal C,\Pi)$, the minimum base length within the two-parent family generated by $(\mathcal C,\Pi)$ and $(\mathcal C_0,\Pi_0)$ is
\[
N_{\mathrm{two\text{-}parent}}(k') = \min_{\substack{ a\geq1,\;b\geq0\\ a,b\in\mathbb Z,\;ak+b=k'}} \left(an+bn_0\right).
\]
For a minimizing pair $(a,b)$, the corresponding bare code is
\[
(\mathcal C,\Pi)^{\odot a} \odot (\mathcal C_0,\Pi_0)^{\odot b},
\]
with the second factor omitted when $b=0$.
Any larger physical length is obtained by applying the required number of pendant extensions.

\noindent These are closure results.
The pendant extension maintains the minimum distance at three and introduces the weight-two stabilizers $Z_hX_\ell$.
Blockwise composition forms a direct product of independent code blocks.
Together, the two constructions establish existence and preserve the BAC condition under the stated fault model.
They do not address circuit-depth or pseudo-threshold optimization.
In this work, the explicit $[[6,1,3]]_{\mathrm{BAC}}$  for $\mathcal M_{\mathrm a}$ and the $[[7,1,3]]_{\mathrm{BAC}}$ for $\mathcal M_{\mathrm c}$  which is used as parent code are graph codes. 
This motivates the graph-message representation introduced next, which provides the stabilizer, syndrome, and logical information needed for the stabilizer codes and their ordering search.

\color{black}
\section{GRAPH CODE CONSTRUCTION USING MEASUREMENT}
\label{sec: MBQC construct QECC}

\noindent This section explains the construction of graph codes using Measurement-based quantum computing~\cite{PhysRevA.68.022312}. 
Measurement-based quantum computation uses an entangled resource state and a sequence of single-qubit measurements to process quantum information~\cite{patil2023clifford} .
A graph state is specified by a simple undirected graph \(\mathsf G=(\mathcal V_G,\mathcal E_G)\).
Each vertex in \(\mathcal V_G\) represents a qubit initialized in \(|+\rangle\), and each edge in \(\mathcal E_G\) represents a controlled-\(Z\) interaction as seen in Fig.~\ref{fig: 2D cluster state}.

\begin{figure}[ht]
    \centering
    \begin{tikzpicture}[node distance={10mm}, thick, main/.style = {draw, circle, minimum size=0.4cm}]
\node[main] (1) {$|{+}\rangle$}; 
\node[main] (2) [ right of=1] {};
\node[main] (3) [right of = 2] {}; 
\node[main] (4) [right of= 3] {};
\node[main] (5) [below of = 1] {}; 
\node[main] (6) [below of= 2] {};
\node[main] (7) [below of = 3] {}; 
\node[main] (8) [below of = 4] {}; 
\node[main] (9) [below of = 5] {}; 
\node[main] (10) [below of = 6] {}; 
\node[main] (11) [below of = 7] {$\mathcal V_G$}; 
\node[main] (12) [below of = 8] {}; 
\draw (1) -- node[midway, above right, sloped ] {} (2);
\draw (2) -- node[midway, above right, sloped ] {} (3);
\draw (3) -- node[midway, above , sloped ] {$CZ$} (4);
\draw (5) -- node[midway, above right, sloped ] {} (6);
\draw (6) -- node[midway, above right, sloped ] {} (7);
\draw (7) -- node[midway, above right, sloped ] {} (8);
\draw (9) -- node[midway, above right, sloped ] {} (10);
\draw (10) -- node[midway, above right, sloped ] {} (11);
\draw (11) -- node[midway, above, sloped ] {$\mathcal E_G$} (12);
\draw (1) -- node[midway, above right, sloped ] {} (5);
\draw (5) -- node[midway, above right, sloped ] {} (9);
\draw (2) -- node[midway, above right, sloped ] {} (6);
\draw (6) -- node[midway, above right, sloped ] {} (10);
\draw (3) -- node[midway, above right, sloped ] {} (7);
\draw (7) -- node[midway, above right, sloped ] {} (11);
\draw (4) -- node[midway, above right, sloped ] {} (8);
\draw (8) -- node[midway, above right, sloped ] {} (12);
\end{tikzpicture}
 \caption{A cluster state represented by a simple undirected graph \(\mathsf G=(\mathcal V_G,\mathcal E_G)\). Vertices represent qubits initialized in \(|+\rangle\), and edges represent controlled-\(Z\) gates.}   
 \label{fig: 2D cluster state}
\end{figure}

\noindent In order to move from the description of graph states to the construction of QECCs, it is important to recall how the measurement outcomes affect the stabilizer structure of a graph state.
The evolution of stabilizers under specific measurement choices provides the foundation for relating cluster states to error-correcting codes.
A formal derivation of the stabilizer generators of post-measurement states can be found in Ref. \cite{nielsen2010quantum}, while the detailed behavior of the stabilizer group under various measurement settings is described in Ref. \cite{patil2023clifford,PhysRevA.69.062311}.
This perspective naturally motivates the construction of quantum error-correcting codes from graph states, known as graph codes.
\noindent Schlingemann and Werner introduced graph-based quantum codes using a graph and a finite Abelian group \cite{Schlingemann_2001}.
Grassl, Klappenecker, and Rotteler, and independently Schlingemann, established the equivalence between stabilizer codes and graph-code descriptions ~\cite{1023317,schlingemann2001stabilizercodesrealizedgraph}.
A measurement-based encoding can be obtained by coupling \(k\) message qubits to \(n\) data qubits with controlled-\(Z\) gates and then measuring the message qubits in the \(X\) basis ~\cite{hwang2015relationgraphcodegraph}.

\noindent Let \(A^{cc}\in\mathbb F_2^{n\times n}\) be the adjacency matrix of the data graph and let \(A^{cm}\in\mathbb F_2^{n\times k}\) describe the edges from the data qubits to the message qubits.
Before the message measurements, the joint system has the \(n\) independent stabilizers
\[
\widetilde g_j
 =X_j\prod_{i=1}^{n}Z_i^{A^{cc}_{ji}}
       \prod_{\mu=1}^{k}Z_{m_\mu}^{A^{cm}_{j\mu}},
 \qquad j=1,\ldots,n.
\]
Their parity check matrix is $\bigl[I_n\;0_{n\times k}\mid A^{cc}\;A^{cm}\bigr]$ ~\cite{8269075}.
If \(A^{cm}\) has column rank \(k\), elimination of one pivot row for each \(X\)-measured message qubit leaves \(n-k\) independent stabilizers with operators applied on data only and hence an \([[n,k]]\) stabilizer code. Algorithm~\ref{algo: matrix method} in Appendix:\ref{Appendix: matrix method for encoding} explains how to encode using the matrix method.

\noindent The graph code in this work uses one message qubit.
In this case let \(b\in\mathbb F_2^n\setminus\{0\}\) be the sole column of \(A^{cm}\), and
choose a pivot \(p\) with \(b_p=1\).
The message column gives the logical Z
\begin{equation}
  \overline Z=\prod_{j=1}^{n}Z_j^{b_j},  
  \label{eq:logicalZgraph}
\end{equation}
while the pivot row gives logical X
\begin{equation}
    \overline X=X_p\prod_{j=1}^{n}Z_j^{A^{cc}_{pj}}.
    \label{eq:logicalXgraph}
\end{equation}

\noindent Since \(\operatorname{wt}(\overline Z)=\operatorname{wt}(b)\), the code distance satisfies \(d\leq\operatorname{wt}(b)\) with proof in Appendix.\ref{APPENDIX: MBQC min distance}.
Therefore, \(\operatorname{wt}(b)\geq d\) is a necessary condition for distance \(d\), but it is not sufficient.

\noindent In the next section, we analyze the graph code parameters responsible for the bare codes.
All analysis is conducted through the lens of linear algebra, and we compare the results of $[[7,1,3]]$ bare codes with $[[8,1,3]]$ bare codes constructed using graph codes.
\color{black}

\section{Graph-theoretic criteria for bare-ancilla implementation of graph codes}
\label{sec:graph-theoretic-bac}

\noindent Earlier sections showed that BACs can be constructed under both \(\mathcal M_{\mathrm a}\) and \(\mathcal M_{\mathrm c}\) fault models.
In this section, we consider \(\mathcal M_{\mathrm c}\) fault model because the errors generated due to \(\mathcal M_{\mathrm a}\) are already in the error set generated because of \(\mathcal M_{\mathrm c}\).
Thus, any circuit that is valid under \(\mathcal M_{\mathrm c}\) is also valid under \(\mathcal M_{\mathrm a}\).
We will prove in the section~\ref{subsec:graph-error-coordinates} that bare code construction is not solely dependent on graphs and show that various components are involved in the hook error.
In the end, we compare the results for the BAC datasets $[[7,1,3]]$ and $[[8,1,3]].$  

\noindent For one message qubit, define
\[
A:=A^{cc}\in\mathbb F_2^{n\times n},
\qquad
b:=A^{cm}\in\mathbb F_2^n\setminus\{0\}.
\]
Thus, \(b_v=1\) exactly when the message qubit is adjacent to the data qubit vertex \(v\in\mathcal V_G\).
For \(x,z\in\mathbb F_2^n\), define the corresponding element of
\(\mathcal P_n\) as
\begin{equation}
    P(x,z):=\bigotimes_{v\in\mathcal V_G}X_v^{x_v}Z_v^{z_v}.
    \label{eq:alzebraic Pauli operator}
\end{equation}

\subsection{Graph code independent of pivot}
\label{subsec:graph-message-stabilizer-space}

\noindent 

\noindent The generators of a cluster state or a graph state using Eq.~\eqref{eq:alzebraic Pauli operator} are  
\begin{equation*}
g_i=P(e_i,Ae_i),
\qquad i\in \mathcal V_G.
\end{equation*}
where \(e_i \) be the binary unit vector at vertex \(``i"\) i.e., $= [0,\cdots,1(at \ i^{th}\  \text{position}),\cdots,0] \in \mathbb F_2^n$.
When the graph state is associated with a message qubit, then the stabilizer generators corresponding to the graph vertex $i$ is $g_iZ_m^{b_i}$.
Thus, it contains $Z_m$ exactly when $b_i=1$, meaning that vertex $i$ is connected to the message qubit~\cite{Schlingemann_2001,basak2026fasteroptimaldecodergraph}.
A pivot is any vertex $p\in\operatorname{supp}(b)$.
Graph code has different stabilizer generators depending on the pivot we choose, but the whole stabilizer group is always the same. We will now prove this using the below lemma.
\begin{lemma}
\label{lem:pivot-independent-stabilizer-group}
For every pivot $p\in\operatorname{supp}(b)$, the set
\begin{equation*}
 \mathcal S : = \{g_i:b_i=0\} \cup \{g_ig_p:b_i=1,\ i\neq p\}
\label{eq:canonical-pivot-basis}
\end{equation*}
defines the stabilizer generating set for the graph code when the message qubit is measured in the $X$ basis.
\end{lemma}

\begin{proof}
Based on the algorithm~\ref{algo: matrix method} in Appendix:\ref{Appendix: matrix method for encoding} after the measurement of the message qubit, the modified stabilizer generator is $g_ig_p^{b_i}$ for $i\neq p$.
As every $g_i$ is defined with the help of unit vector $e_i$ therefore the  vector of this modified generator $g_ig_p^{b_i}$ is
\begin{equation}
t_i^{(p)}=e_i+b_i e_p, \qquad i\neq p.
\label{eq:t_i defined}
\end{equation}
It satisfies $b^Tt_i^{(p)}=0$, and hence lies in
\begin{equation*}
b^\perp=\{u\in\mathbb F_2^n:b^Tu=0\}.
\end{equation*}
The $n-1$ vectors $t_i^{(p)}$ are independent because, for each $i\neq p$, only $t_i^{(p)}$ has a nonzero $i$th coordinate.
Since $\dim b^\perp=n-1$, they form a basis of $b^\perp$.
% And we see that elements of $\mathcal {B} $ are independent of $p$, which are the generators of the graph state. 
Therefore,  $\mathcal S$  is independent of $p$.
\end{proof}

\noindent Different pivots can nevertheless give distinct stabilizer sets, hence different hook-error sets.
Therefore, next, we establish the effect of stabilizer elements on hook errors.

\subsection{Hook error dependency on graph code}
\label{subsec:graph-error-coordinates}

\noindent For any graph code that needs to be bare code, we need to analyze the hook error dependency on the graph code parameters.
% For that, we need to define the various parameters of the graph code in algebraic form. 

% \noindent
% For an Pauli operator $E=P(x,z)$, define\cite{cross2008codeword,chuang2009codeword}
% \begin{equation}
% \begin{aligned}
% & q_A(E)=z+Ax,
% \quad 
% Q_b=\mathbb F_2^n/\langle b\rangle,
% \\
% & \sigma_{A,b}(E)=[q_A(E)]\in Q_b. \\ &
% \Phi_{A,b}(E)=\bigl(q_A(E),b^Tx\bigr)
% \in\mathbb F_2^n\times\mathbb F_2.
% \end{aligned}
% \label{eq:full-graph-coset-coordinate}
% \end{equation}
% where $q_A(E)$ is an $n$-bit vector that depicts the operator $E$ into a purely $Z$ basis using graph stabilizer generators $g_i$, also called the graph-coordinate vector.
% $Q_b$, called as graph-coordinate class, is the quotient space in which two n-bit vectors $q$ and $q+b$ are treated as equal; it is also equivalent to the syndrome space, $i.e \ 2^{n-1}$.
% The reason for defining this quantity is that any  $q_A(E)$ and $q_A(E) + b$ syndrome will be the same, hence $\sigma_{A,b}(E)$ is defined.
% Lastly, $\Phi_{A,b}(E)$ is an $n+1$ vector where the $n$-bit vector corresponds to $E$ with another 1 bit, which captures the commutation relation with the logical $Z$ operator.
% Also we retain the single qubit error set \(\mathcal E_1\) defined in Sec.~\ref{subsec:exact-bac-condition}.

\noindent Let \(\mathcal C(A,b)\) denote the \([[n,1]]\) code stabilized by \(\mathcal S\).
With  $c\in\mathbb F_2^n$ such that $b^Tc=1$ then using the Eq.~\eqref{eq:t_i defined} the logical operators are $\overline Z=P(0,b)$ and $\overline X=P(c,Ac)$.
We know that stabilizer generators using  \(u\in b^\perp\setminus\{0\}\) can be written as $g_u = P(u,Au)$.

\noindent The support of $g_u$ is
\begin{equation}
W(u):=\operatorname{supp}(u)\cup\operatorname{supp}(Au).
\end{equation}
For each $v\in W(u)$, define the corresponding single-qubit Pauli operator by
\begin{equation}
\mu_v(u)
:=
P\bigl(u_ve_v,(Au)_ve_v\bigr).
\end{equation}
The two bits $u_v$ and $(Au)_v$ determine the Pauli operator acting at vertex $v$.
Therefore,
\begin{equation}
g_u
=
\prod_{v\in W(u)}\mu_v(u).
\end{equation}

\noindent For an Pauli operator $E=P(x,z)$  define 
\[ 
q_A(E)=z+Ax \in \mathbb F^n_2,
\]
is an $n$-bit vector that depicts the operator $E$ into a purely $Z$ basis using graph stabilizer generators $g_i$ \cite{crosscodeword,chuang2009codeword}.

\noindent Hence $\mu_v(u)$ can be written as  
\begin{align}
\left(
(Au)_ve_v+u_vAe_v,
\right).
\label{eq:local-graph-contribution}
\end{align}

\noindent With \(\pi=(v_1,\ldots,v_w)\) be an interaction order for the stabilizer generator \(g_u\), the hook error due to a fault at location \(t\) in the ancilla is 
\begin{equation}
 E_{hook}
 =\left(\prod_{j<t}\mu_{v_j}(u)\right)E_{v_t},
\qquad
t=2,\ldots,w-1.
\label{eq:graph-to-circuit-hook-relation}
\end{equation}
where \(E\in\{X,Y,Z\}\).

\noindent Eq.~\eqref{eq:graph-to-circuit-hook-relation} introduces the effect of every component regarding the hook error.
From Eq.\ref{eq:local-graph-contribution} $(A,b)$ and stabilizer vector $u$ in Eq.~\eqref{eq:graph-to-circuit-hook-relation} determine each contribution and interaction order $\pi$ determines the order in which these contributions are added, whereas fault location $t$ determines where the product stops. 
It shows that various factors contribute to hook error generation; the graph property alone is not responsible for hook errors. 
Now we analyse the $[[7,1,3]]$ and $[[8,1,3]]$ graph codes for the error model $\mathcal M_a$.

\subsection{Analysis of the graph codes: $[[7,1,3]]$ and $[[8,1,3]]$ BACs}
\label{subsec:joint-ordering-bare-rank}

\noindent Fix a stabilizer generator set $\mathcal S=\{g_1,\ldots,g_{n-1})\}$.
For each generator $g_j$, let $\mathtt{ Ord\_list}_j $ be the set that contains the valid permutations of the operators which will be obtained using Algorithm~\ref{alg:local-order-search}.
\noindent Now, to analyze the graph codes, we use Algorithm~\ref{alg:global-basis-ordering-search} to find the bare codes. 
All nonisomorphic connected simple parent graphs used in this analysis are taken from the McKay catalog~\cite{McKayGraphData}.
The correlated fault model $\mathcal M_c$ of Sec.~\ref{sec:correlated error two qubit gate} is used throughout.

\noindent The seven and eight qubit datasets are analyzed using the three stage search as shown in Fig.~\ref{fig:three-stage-basis-search}.
Each successive stage enlarges the set of stabilizer generators under consideration.
Stage~1 examines the stabilizer generators obtained from message-qubit elimination.
For each graph-message pair $(A,b)$ and every pivot $p\in\operatorname{supp}(b)$, all allowed interaction orders are tested for the corresponding generators $\mathcal S$.
A case with valid code is considered and is resolved at Stage~1.
Cases for which every canonical pivot fails are passed to Stage~2, where it searches for non-canonical stabilizer generators from $\mathcal S$.
The first search applies depth-one operations of the form
\begin{equation*}
g_i\longmapsto g_i g_j,
\qquad i\neq j,
\label{eq:depth-one-row-repair}
\end{equation*}
to the canonical generators.
If no witness is found, a randomized search assembles independent stabilizer generators from the stabilizer set that have at least one locally valid interaction order.
For further unresolved cases, Stage~3 performs an exhaustive search over all possible stabilizer generators.
If the exhaustive search finds no such generators, the case is classified as exactly nonbare under $M_c$.

\begin{figure}[t]
\centering
\begin{tikzpicture}[scale=0.95,
  node distance=0.62cm and 0.85cm,
  stage/.style={
    rounded corners=2.5pt,
    draw=neutralGray,
    fill=white,
    line width=0.7pt,
    text width=4.70cm,
    align=left,
    inner sep=6pt,
    font=\footnotesize
  },
  start/.style={
    rounded corners=2.5pt,
    draw=searchBlue,
    fill=searchFill,
    line width=0.6pt,
    text width=3.65cm,
    align=center,
    inner sep=5pt,
    font=\footnotesize
  },
  witness/.style={
    rounded corners=2.5pt,
    draw=witnessGreen,
    fill=witnessFill,
    line width=0.7pt,
    text width=1.68cm,
    align=center,
    inner sep=5pt,
    font=\scriptsize
  },
  nonbare/.style={
    rounded corners=2.5pt,
    draw=nonbareRed,
    fill=nonbareFill,
    line width=0.7pt,
    text width=2.85cm,
    align=center,
    inner sep=5pt,
    font=\scriptsize
  },
  flow/.style={
    -{Latex[length=2.0mm,width=1.4mm]},
    draw=neutralGray,
    line width=0.65pt
  },
  branch/.style={
    font=\scriptsize,
    text=neutralGray,
    fill=white,
    inner sep=1.2pt,
    align=center
  }
]

\node[start] (input) {
  Graph-message case $(A,b)$
};

\node[stage,below=of input] (stage1) {
  \textbf{Stage 1: Canonical-pivot search}\\[-1pt]
  Test every canonical generator set $\mathcal S$,
  $p\in\operatorname{supp}(b)$.\\[2pt]
  
};

\node[witness,right=of stage1] (witness1) {
  Canonical\\generator BAC\\[2pt]
  \textit{at Stage 1}
};

\node[stage,below=0.92cm of stage1] (stage2) {
  \textbf{Stage 2: Constructive generator changes}\\[-1pt]
  Try depth-one generators $g_i\mapsto g_i g_j$ and randomized independent
  generators. 
};

\node[witness,right=of stage2] (witness2) {
  Noncanonical\\generator BAC\\[2pt]
  \textit{at Stage 2}
};

\node[stage,below=0.92cm of stage2] (stage3) {
  \textbf{Stage 3: All-generator search}\\[-1pt]
  Test every full-rank generator set from
  $\mathcal S$.\\[2pt]
  
};

\node[witness,right=of stage3] (witness3) {
  All-generator\\witness\\[2pt]
  \textit{resolved at Stage 3}
};

\node[nonbare,below=0.72cm of stage3] (failure) {
  Exactly nonbare under $\mathcal M_{\mathrm c}$
};

\draw[flow] (input) -- (stage1);

\draw[flow] (stage1.east) --
  node[branch,above] {\textcolor{black}{passes}}
  (witness1.west);
\draw[flow] (stage1.south) --
  node[branch,right] {\textcolor{black}{all canonical}\\ \textcolor{black}{generators fail}}
  (stage2.north);

\draw[flow] (stage2.east) --
  node[branch,above] {\textcolor{black}{passes}}
  (witness2.west);
\draw[flow] (stage2.south) --
  node[branch,right] {\textcolor{black}{no witness}\\ \textcolor{black}{inconclusive}}
  (stage3.north);

\draw[flow] (stage3.east) --
  node[branch,above] {\textcolor{black}{passes}}
  (witness3.west);
\draw[flow] (stage3.south) --
  node[branch,right] {\textcolor{black}{failed}}
  (failure.north);

\end{tikzpicture}
\caption{Three-stage generator-and-ordering search for a graph-message pair $(A,b)$.
Stage~1 exhausts the canonical pivot generators, Stage~2 tests selected noncanonical generators, and Stage~3 performs search from the whole stabilizer group.
Only an exhaustive failure at Stage~3 proves that the case is nonbare under $\mathcal M_{\mathrm c}$.}
\label{fig:three-stage-basis-search}
\end{figure}

\subsubsection{Analysis of the  $[[7,1,3]]$ graph code dataset}
\label{subsubsec:enumerated-seven-qubit-results}

\noindent
The seven-qubit enumeration contains the $853$ graphs~\cite{McKayGraphData}.
For each parent adjacency matrix $A$, every message-neighborhood vector $b\in\mathbb F_2^7$ satisfying $3\leq\operatorname{wt}(b)\leq7$ is included.
The number of such vectors for each parent is $\sum_{w=3}^{7}\binom{7}{w}=99.$
Hence, the total graph-message candidates $(A,b)$ domain are $ N_{\mathrm{cand}}^{(7)} = 84\,447$.
The exact distance test classifies $8\,806$ candidates as distance one, $72\,262$ as distance two, and $3\,379$ as distance three.

\noindent Each distance-three pair $(A,b)$ is counted as one graph-message \emph{case}.
For each case, every vertex $p\in\operatorname{supp}(b)$ can be selected as the pivot for the graph code.
Each pivot choice determines a potentially distinct set of canonical stabilizer generators and is stored as a single \emph{pivot record}.
\noindent Out of $3\,379$ cases, $112$ cases have BAC for every pivot, $673$ cases have pivots that yield both BAC and non-BAC, and the remaining $2\,594$ cases have no canonical generators for any pivot that satisfy the BAC properties. 
Therefore, $785$ cases have at least one canonical pivot BAC.
Only the $2\,594$ Stage~1 failures enter Stage~2.
The depth-one search resolves $1\,194$ cases, and the randomized independent-generator search resolves a further $1\,108$.
The remaining $292$ cases enter the exact all-generator search.
Stage~3 finds six independent stabilizer generators for $90$ cases and proves that no such stabilizer generator set exists for the other $202$ under $M_c$.
All $3\,177$  are valid BACs.
The $202$ negative cases exhibit two different obstructions.
For $158$ cases, stabilizers generator sets have no locally valid independent generators.
The remaining $44$ cases have locally valid six generators, but fail to resolve the hook error set as combined.
The final classification is
\begin{equation*}
N_{\mathrm{case}}^{(7)}=3\,379, \qquad
N_{\mathrm{BAC}}^{(7)}=3\,177, \qquad
N_{\mathrm{nonBAC}}^{(7)}=202.
\label{eq:enumerated-seven-bare-rank-result}
\end{equation*}
Hence, $94.02\%$ of the distance-three graph-message cases are BACs.
Table~\ref{tab:enumerated-seven-bare-results} summarizes the complete classification.
\begin{table}[ht]
\color{black}
\centering
\resizebox{\columnwidth}{!}{%
\begin{tabular}{lrrr}
\toprule
Classification & Count & Denominator & Fraction \\
\midrule
\multicolumn{4}{l}{\textit{Stage 1: canonical-pivot analysis}}\\
Every canonical pivot passes
& $112$ & $3\,379$ & $3.31\%$ \\
Some but not all canonical pivots pass
& $673$ & $3\,379$ & $19.92\%$ \\
At least one canonical pivot passes
& $785$ & $3\,379$ & $23.23\%$ \\
No canonical-pivot witness
& $2\,594$ & $3\,379$ & $76.77\%$ \\
\midrule
\multicolumn{4}{l}{\textit{Stage 2: noncanonical generator search}}\\
Depth-one  witness
& $1\,194$ & $2\,594$ & $46.03\%$ \\
Randomized independent-generator witness
& $1\,108$ & $2\,594$ & $42.71\%$ \\
Unresolved after Stage~2
& $292$ & $2\,594$ & $11.26\%$ \\
\midrule
\multicolumn{4}{l}{\textit{Stage 3: exact all-generator search}}\\
Exact valid witness
& $90$ & $292$ & $30.82\%$ \\
Exact nonbare code
& $202$ & $292$ & $69.18\%$ \\
\midrule
Final valid BAC
& $3\,177$ & $3\,379$ & $94.02\%$ \\
\bottomrule
\end{tabular}%
}
\caption{Bare-ancilla classification of the connected-parent $[[7,1,3]]$ graph-message dataset.
All rows are reported at the graph-message level.
The Stage~2 fractions are measured relative to the $2\,594$ cases entering that stage, while the Stage~3 fractions are measured relative to the $292$ cases unresolved after Stage~2.}
\label{tab:enumerated-seven-bare-results}
\end{table}

\vspace{-0.5cm}
\subsubsection{Analysis of the $[[8,1,3]]$ graph code dataset}
\label{subsec:enumerated-eight-qubit-results}

\noindent The eight-qubit analysis  uses all $11\,117$ nonisomorphic graphs~\cite{McKayGraphData}.
For each adjacency matrix of the parent graph, every message-neighborhood vector $b\in\mathbb F_2^8$ satisfying $3\leq\operatorname{wt}(b)\leq8$ is included.
Since $\sum_{w=3}^{8}\binom{8}{w}=219,$ the total  graph-message pair candidates are $N_{\mathrm{cand}}^{(8)}=2\,434\,623$ where $145\,299$ of distance one, $1\,953\,909$ of distance two, and $335\,415$ of distance three.
For valid $N_{\mathrm{case}}^{(8)}= 335\,415,$ cases, all canonical pivots pass for $301\,304$ cases, while $20\,745$ cases passes for few pivots.
Consequently, at least one canonical pivot works for $322\,049$ of the $335\,415$ cases.
Stage~2 is applied to the remaining $13\,366$ cases.
A depth-one row operation gives BAC for $11\,291$ cases, and the randomized independent-generator search finds a witness for the remaining $2\,075$ cases.
The final classification is therefore
\begin{equation}
N_{\mathrm{case}}^{(8)} = N_{\mathrm{BAC}}^{(8)} = 335\,415,
\qquad
N_{\mathrm{nonBAC}}^{(8)} = 0.
\label{eq:enumerated-eight-bare-rank-result}
\end{equation}
Hence, every eight-qubit graph-message case has a valid set of bare-ancilla stabilizer generators under $M_c$.
The classification is summarized in Table~\ref{tab:enumerated-eight-bare-results}.\\ 

\begin{table}[t]
\color{black}
\centering
\resizebox{\columnwidth}{!}{%
\begin{tabular}{lrrr}
\toprule
Classification & Count & Denominator & Fraction \\
\midrule
\multicolumn{4}{l}{\emph{Stage~1: canonical-pivot analysis}}\\
Every canonical pivot passes
& $301\,304$ & $335\,415$ & $89.83\%$ \\
Some but not all canonical pivots pass
& $20\,745$ & $335\,415$ & $6.18\%$ \\
At least one canonical pivot passes
& $322\,049$ & $335\,415$ & $96.02\%$ \\
No canonical-pivot witness
& $13\,366$ & $335\,415$ & $3.98\%$ \\
\midrule
\multicolumn{4}{l}{\emph{Stage~2: noncanonical generator search}}\\
Depth-one row-repair witness
& $11\,291$ & $13\,366$ & $84.48\%$ \\
Randomized independent-generator witness
& $2\,075$ & $13\,366$ & $15.52\%$ \\
\midrule
Cases entering the exact Stage~3 search
& $0$ & $13\,366$ & $0.00\%$ \\
\midrule
Final valid BAC 
& $335\,415$ & $335\,415$ & $100.00\%$ \\
\bottomrule
\end{tabular}%
}
\caption{Bare-ancilla classification of the connected-parent $[[8,1,3]]$ graph-message dataset.
The Stage~1 quantities are evaluated at the graph-message level.
The Stage~2 and Stage~3 denominators contain only the $13\,366$ cases with no canonical-pivot witness.}
\label{tab:enumerated-eight-bare-results}
\end{table}

\noindent Both datasets show that the choice of stabilizer generators and interaction orders is important for a valid BAC to the corresponding fault model. 
However, the set of canonical stabilizer generators is much more effective for the eight-qubit case.
At stage~1, at least one canonical pivot gives BAC for $23.23\%$  of $[[7,1,3]]$ cases and $96.02\%$  of $[[8,1,3]]$ cases.
The full-stage search gives $94.02\%$ BAC for the seven-qubit dataset and $100\%$ for the eight-qubit dataset.
Stage~3 is required for $292$ seven-qubit cases but zero eight-qubit cases.
The seven-qubit codes have 64 possible syndrome spaces, while the eight-qubit codes have 128.
The larger syndrome space provides more capacity to distinguish hook errors, but its size does not determine whether a BAC exists.

\noindent The parent graph alone also does not determine the final classification.
Of the $853$ seven-vertex parent graphs, $666$ gives the $[[7,1,3]]$ graph codes for at least one graph-message case, as each parent graph is paired with \(99\) allowed message vectors.
Among these $666$, $464$ support only bare cases, $157$ support only nonbare cases, and $45$ support both as the message vector $b$ changes.
For a fixed pair $(A,b)$, its Stage~1 result can also depend on the selected canonical pivot.
However, the final bare or nonbare classification is determined by the graph-message pair after optimization over the set of allowed stabilizer generators and interaction orders.
These results determine whether a valid BAC exists, but they do not measure how well that code suppresses logical errors.
The next section, therefore, compares the logical performance of selected seven- and eight-qubit implementations under circuit-level noise.

\section{Simulation setup and logical performance analysis}
\label{sec:numerical-graph-performance}

\noindent  This section compares implementations of the $[[7,1,3]]$ and $[[8,1,3]]$ graph codes.
Variations in logical performance with respect to the graph structure, stabilizer generators, interaction ordering, extraction method and noise model are to be examined, with different block lengths treated as an explicit part of the comparison.

\subsection{Simulation protocol and decoder}
\label{subsec:simulation-protocol-final}

\noindent The stabilizer circuits are simulated using the CHP (CNOT-Hadamard-Phase) framework~\cite{PhysRevA.70.052328}.
Each Monte Carlo trial contains a noisy error-correction block followed by a final ideal error-correction block.
During the noisy block, stabilizer generators are measured sequentially.
If any individual stabilizer measurement returns a nonzero outcome, all stabilizers are measured again, and a correction is applied using the decoder.
The final ideal error-correction block is performed regardless of the outcomes obtained during the noisy block. After this ideal correction, final correction identifies whether the propagated error belongs to the stabilizer group or to a nontrivial logical coset.
A trial is counted as a logical failure only when the syndrome is zero and the remaining error corresponds to a nontrivial logical operator. 
\\

\noindent 
The BAC condition establishes that the specified hook errors can be corrected when a corresponding syndrome is available.
However, a fault occurring late during noisy syndrome extraction sequence may introduce a data error that is not detected.
In our simulations, the subsequent ideal error-correction block can detect and correct such an error, provided the decoder assigns an appropriate correction operator.
Therefore, the reported logical-error rates and pseudo-thresholds evaluate the performance of the implemented noisy error-correction circuit followed by ideal recovery, rather than establishing fault tolerance of the complete noisy error-correction procedure.
\\

% If the initial syndrome measurement returns a nonzero syndrome, all stabilizers are measured again, and a correction is applied using a decoder.
% The final correction identifies whether the propagated error belongs to the stabilizer group or to a nontrivial logical coset.
\noindent The decoder first assigns the syndromes to the single-qubit errors.
It next assigns the syndromes for hook errors stored for faults in the syndrome-extraction circuit.
Any remaining syndrome is passed to the Look-Up-Table(LUT) decoder.
A trial is counted as a logical failure when the final error commutes with every stabilizer generator but belongs to a nontrivial logical coset.
For total trials $N_{\mathrm{T}}$,
\begin{equation*}
 P_{\mathrm{L}}(p)
 = \frac{N_{\mathrm{L}}(p)}{N_{\mathrm{T}}},
 \label{eq:logical-error-rate-final}
\end{equation*}
where $N_{\mathrm{L}}(p)$ is the number of logical failures.

\noindent Two circuit-level noise models are used.
\noindent Under depolarizing noise, a single-qubit gate is followed by $X$, $Y$, or $Z$ with probability $p_s/3$ each.
A two-qubit gate is followed by one of the $15$ nonidentity two-qubit Pauli operators with probability $p_t/15$ each.
A measurement outcome is followed by a bit flip error with probability $p_m$, while preparation of $\lvert0\rangle$ or $\lvert+\rangle$ fails with probability $p_p$.

\noindent The anisotropic model represents gate-aligned faults associated with under- or over-rotation in trapped-ion gates ~\cite{parrado2020crosstalk,PhysRevLett.82.1971}.
A controlled-$P$ interaction is followed by $Z\otimes P$ with probability $p_t$.
The interaction qubits also undergo single-qubit depolarizing noise.
For the simulation, the parameters are defined as $p=p_t$, with $p_s=p_m=p_p=4p/15$.
The same rates apply to the bare and flag circuits.\\

\noindent For each implementation, $20$ logarithmically spaced values in $10^{-4}\leq p\leq10^{-2}$ are simulated with $N_{\mathrm{T}}=10^6$ trials per point.
The pseudo-threshold is defined by
\begin{equation}
 P_{\mathrm{L}}(p_{\mathrm{th}})=p_{\mathrm{th}},
 \label{eq:pseudothreshold-final}
\end{equation}
and is obtained by log-log interpolation between the two adjacent points that bracket the crossing.
\noindent From the complete bare-code datasets, we selected five hundred \([[7,1,3]]\) records and one thousand \([[8,1,3]]\) records.
The samples included valid BAC records by the canonical-pivot, depth-one generators $g_i'=g_ig_j$ for $i\neq j$, as well as randomized stabilizer generator search.
Table~\ref{tab:ensemble-summary-final} summarizes the two enumeration pipelines and the simulated samples.

\begin{table}[t]
\centering
\color{black}
\begin{ruledtabular}
\begin{tabular}{lrr}
Quantity & $[[7,1,3]]$ & $[[8,1,3]]$ \\
\hline
Parent graphs          & $853$      & $11\,117$ \\
Distance-three cases   & $3\,379$   & $335\,415$ \\
Canonical successes    & $785$      & $322\,049$ \\
Final bare cases       & $3\,177$   & $335\,415$ \\
Simulated records      & $500$      & $1\,000$ \\
Pseudo-threshold crossings & $500$  & $1000$ \\
\end{tabular}
\end{ruledtabular}
\caption{Enumeration and simulation samples. A canonical success means
that at least one canonical message pivot passes without changing the
stabilizer generators.}
\label{tab:ensemble-summary-final}

\end{table}

\subsection{Logical-error rates and pseudo-threshold analysis}
\label{subsec:logical-error-threshold-analysis}

\noindent  Each bare code fixes the message-neighborhood set $b$, pivot $p$, stabilizer generator set $\mathcal S$, and interaction order $\pi$, in addition to the parent graph $G$.
Logical performance further depends on the extraction method $m$, noise model $\mathcal M$, and decoder $D$. 
The appropriate relation is therefore
\begin{equation*}
 p_{\mathrm{th}}=f(G,b,p,\mathcal S,\pi;m,\mathcal M,D),
\end{equation*}
rather than a graph-only map.

\noindent For a measured generator $g$, let $n_X(g)$, $n_Y(g)$, and $n_Z(g)$ count its Pauli operators.
An $X$ or $Z$ operator uses one data-ancilla two-qubit interaction, while a $Y$ operator is implemented by a controlled-$Z$ followed by a controlled-$X$.
We define the effective interaction count
\begin{equation*}
\begin{aligned}
 Q
 &=\sum_{g\in\mathcal{B}}
 [n_X(g)+2n_Y(g)+n_Z(g)]\\
 &=\sum_{g\in\mathcal{B}}[w(g)+n_Y(g)].
\end{aligned}
\label{eq:effective-interaction-count-final}
\end{equation*}
The seven-qubit sample spans $24\leq Q\leq47$, with median $Q=32$, while the eight-qubit sample spans $25\leq Q\leq50$, with median $Q=38$.
Dividing the seven-qubit records into low ($24\leq Q\leq30$), intermediate ($31\leq Q\leq34$), and high ($35\leq Q\leq47$) groups gives $168$, $190$, and $142$ records, respectively.
The median bare pseudo-threshold decreases from $1.027\times10^{-3}$ in the low-$Q$ group to $0.519\times10^{-3}$ in the high-$Q$ group under anisotropic noise.
Under depolarizing noise, it decreases from $2.499\times10^{-3}$ to $1.321\times10^{-3}$.
Figure~\ref{fig:median-ler-Q-final} shows that the same ordering holds across the simulated logical-error curves.
% Drop-in PGFPlots replacement for the two median-LER PDF panels.
% The 120 plotted values are embedded below, so no external CSV file is needed.
%
% Add the following lines to the manuscript preamble:
%   \usepackage{pgfplots}
%   \pgfplotsset{compat=1.18}
%
% Replace the old figure environment by the code below.

\definecolor{lowQ}{HTML}{1F77B4}
\definecolor{midQ}{HTML}{FF7F0E}
\definecolor{highQ}{HTML}{2CA02C}

\begin{figure}[t]
\centering

\begin{tikzpicture}[scale=0.95]
\begin{loglogaxis}[
  width=0.95\columnwidth,
  height=0.64\columnwidth,
  xmin=1e-4,
  xmax=1e-2,
  xtick={1e-4,1e-3,1e-2},
  ytick={1e-5,1e-4,1e-3,1e-2,1e-1},
  xlabel={Physical error probability $p$},
  ylabel={Median logical-error rate $P_{\mathrm L}$},
  tick label style={font=\scriptsize},
  label style={font=\small},
  title style={font=\small},
  grid=both,
  major grid style={draw=gray!28,line width=0.25pt},
  minor grid style={draw=gray!14,line width=0.20pt},
  legend style={font=\tiny,at={(0.02,0.98)},anchor=north west,
    draw=gray!50,fill=white,fill opacity=0.92,text opacity=1,
    cells={anchor=west},
    inner xsep=2pt,
  inner ysep=1pt,
  row sep=-1pt,
  /tikz/every even column/.append style={column sep=2pt}},
  legend cell align=left,
  every axis plot/.append style={line width=0.85pt},
  clip marker paths=true,
  title={Bare $[[7,1,3]]$ records: anisotropic noise}
]
\addplot+[
  color=lowQ,
  mark=*,
  mark size=1.15pt
] coordinates {
    (0.0001,9e-06)
    (0.0001274274985703,1.6e-05)
    (0.0001623776739188,2.7e-05)
    (0.0002069138081114,4.2e-05)
    (0.000263665089873,6.9e-05)
    (0.0003359818286283,0.00011)
    (0.0004281332398719,0.000181)
    (0.0005455594781168,0.000294)
    (0.0006951927961775,0.0004685)
    (0.00088586679041,0.000761)
    (0.0011288378916846,0.0012385)
    (0.0014384498882876,0.0019815)
    (0.0018329807108324,0.00314)
    (0.0023357214690901,0.00504)
    (0.0029763514416313,0.0079355)
    (0.0037926901907322,0.012634)
    (0.0048329302385717,0.0195745)
    (0.0061584821106602,0.0304415)
    (0.0078475997035146,0.0463745)
    (0.01,0.069778)
};
\addlegendentry{Low $Q$ ($24$--$30$)}

\addplot+[
  color=midQ,
  mark=*,
  mark size=1.15pt
] coordinates {
    (0.0001,1.3e-05)
    (0.0001274274985703,2.1e-05)
    (0.0001623776739188,3.35e-05)
    (0.0002069138081114,5.5e-05)
    (0.000263665089873,9.1e-05)
    (0.0003359818286283,0.000145)
    (0.0004281332398719,0.000236)
    (0.0005455594781168,0.0003825)
    (0.0006951927961775,0.0006105)
    (0.00088586679041,0.0009925)
    (0.0011288378916846,0.001596)
    (0.0014384498882876,0.002537)
    (0.0018329807108324,0.0040415)
    (0.0023357214690901,0.006451)
    (0.0029763514416313,0.010209)
    (0.0037926901907322,0.015931)
    (0.0048329302385717,0.024918)
    (0.0061584821106602,0.0380885)
    (0.0078475997035146,0.0580425)
    (0.01,0.0866205)
};
\addlegendentry{Intermediate $Q$ ($31$--$34$)}

\addplot+[
  color=highQ,
  mark=*,
  mark size=1.15pt
] coordinates {
    (0.0001,2e-05)
    (0.0001274274985703,3.2e-05)
    (0.0001623776739188,5.3e-05)
    (0.0002069138081114,8.55e-05)
    (0.000263665089873,0.000136)
    (0.0003359818286283,0.0002255)
    (0.0004281332398719,0.000355)
    (0.0005455594781168,0.000574)
    (0.0006951927961775,0.000923)
    (0.00088586679041,0.001459)
    (0.0011288378916846,0.002328)
    (0.0014384498882876,0.003787)
    (0.0018329807108324,0.0059155)
    (0.0023357214690901,0.0094435)
    (0.0029763514416313,0.0148635)
    (0.0037926901907322,0.023143)
    (0.0048329302385717,0.035584)
    (0.0061584821106602,0.054232)
    (0.0078475997035146,0.0811055)
    (0.01,0.1181735)
};
\addlegendentry{High $Q$ ($35$--$47$)}

\addplot[
  black,
  densely dotted,
  line width=0.95pt,
  domain=1e-4:1e-2,
  samples=2
] {x};
\addlegendentry{$P_{\mathrm L}=p$}
\end{loglogaxis}
\end{tikzpicture}

\vspace{0.30cm}

\begin{tikzpicture}[scale=0.95]
\begin{loglogaxis}[
  width=0.95\columnwidth,
  height=0.64\columnwidth,
  xmin=1e-4,
  xmax=1e-2,
  xtick={1e-4,1e-3,1e-2},
  ytick={1e-5,1e-4,1e-3,1e-2,1e-1},
  xlabel={Physical error probability $p$},
  ylabel={Median logical-error rate $P_{\mathrm L}$},
  tick label style={font=\scriptsize},
  label style={font=\small},
  title style={font=\small},
  grid=both,
  major grid style={draw=gray!28,line width=0.25pt},
  minor grid style={draw=gray!14,line width=0.20pt},
  legend style={font=\tiny,at={(0.02,0.98)},anchor=north west,
    draw=gray!50,fill=white,fill opacity=0.92,text opacity=1,
    cells={anchor=west},
    inner xsep=2pt,
  inner ysep=1pt,
  row sep=-1pt,
  /tikz/every even column/.append style={column sep=2pt}},
  legend cell align=left,
  every axis plot/.append style={line width=0.85pt},
  clip marker paths=true,
  title={Bare $[[7,1,3]]$ records: depolarizing noise}
]
\addplot+[
  color=lowQ,
  mark=*,
  mark size=1.15pt
] coordinates {
    (0.0001,4e-06)
    (0.0001274274985703,7e-06)
    (0.0001623776739188,1.1e-05)
    (0.0002069138081114,1.8e-05)
    (0.000263665089873,2.8e-05)
    (0.0003359818286283,4.75e-05)
    (0.0004281332398719,7.6e-05)
    (0.0005455594781168,0.0001235)
    (0.0006951927961775,0.000205)
    (0.00088586679041,0.0003245)
    (0.0011288378916846,0.000523)
    (0.0014384498882876,0.000852)
    (0.0018329807108324,0.0013735)
    (0.0023357214690901,0.002188)
    (0.0029763514416313,0.003507)
    (0.0037926901907322,0.0055845)
    (0.0048329302385717,0.0087905)
    (0.0061584821106602,0.0138525)
    (0.0078475997035146,0.021683)
    (0.01,0.033488)
};
\addlegendentry{Low $Q$ ($24$--$30$)}

\addplot+[
  color=midQ,
  mark=*,
  mark size=1.15pt
] coordinates {
    (0.0001,5e-06)
    (0.0001274274985703,9e-06)
    (0.0001623776739188,1.4e-05)
    (0.0002069138081114,2.5e-05)
    (0.000263665089873,3.9e-05)
    (0.0003359818286283,6.3e-05)
    (0.0004281332398719,0.000102)
    (0.0005455594781168,0.000166)
    (0.0006951927961775,0.000266)
    (0.00088586679041,0.00043)
    (0.0011288378916846,0.0006925)
    (0.0014384498882876,0.001113)
    (0.0018329807108324,0.0018035)
    (0.0023357214690901,0.0028695)
    (0.0029763514416313,0.004612)
    (0.0037926901907322,0.007327)
    (0.0048329302385717,0.011541)
    (0.0061584821106602,0.018095)
    (0.0078475997035146,0.028008)
    (0.01,0.043041)
};
\addlegendentry{Intermediate $Q$ ($31$--$34$)}

\addplot+[
  color=highQ,
  mark=*,
  mark size=1.15pt
] coordinates {
    (0.0001,8e-06)
    (0.0001274274985703,1.3e-05)
    (0.0001623776739188,2.1e-05)
    (0.0002069138081114,3.4e-05)
    (0.000263665089873,5.5e-05)
    (0.0003359818286283,9e-05)
    (0.0004281332398719,0.000148)
    (0.0005455594781168,0.0002365)
    (0.0006951927961775,0.0003745)
    (0.00088586679041,0.000604)
    (0.0011288378916846,0.000976)
    (0.0014384498882876,0.0015695)
    (0.0018329807108324,0.002498)
    (0.0023357214690901,0.003979)
    (0.0029763514416313,0.006314)
    (0.0037926901907322,0.010104)
    (0.0048329302385717,0.015831)
    (0.0061584821106602,0.024591)
    (0.0078475997035146,0.0378885)
    (0.01,0.057512)
};
\addlegendentry{High $Q$ ($35$--$47$)}

\addplot[
  black,
  densely dotted,
  line width=0.95pt,
  domain=1e-4:1e-2,
  samples=2
] {x};
\addlegendentry{$P_{\mathrm L}=p$}
\end{loglogaxis}
\end{tikzpicture}

\caption{Median bare-method logical-error curves for the $[[7,1,3]]$
records grouped by the effective interaction count $Q$.  The upper and lower
panels show anisotropic and circuit-level depolarizing noise, respectively.
The dotted line indicates $P_{\mathrm L}=p$.}
\label{fig:median-ler-Q-final}
\end{figure}

\noindent At low physical error rate, the curves are close to quadratic,
\begin{equation}
 P_{\mathrm{L}}(p)=A_2p^2+O(p^3).
 \label{eq:quadratic-ler-final}
\end{equation}
The median fitted exponent ranges from $1.963$ to $1.971$ across code lengths, extraction methods, and noise models.
Because all allowed single data errors are correctable, this behavior is consistent with pairs of faults providing the leading low-noise contribution.
Solving $P_{\mathrm{L}}=p$ gives $p_{\mathrm{th}}\simeq A_2^{-1}$.

\noindent Within each family, method, and noise model, the pseudo-threshold approximately follows
\begin{equation}
 p_{\mathrm{th}}\simeq C Q^{-\alpha}.
 \label{eq:power-law-final}
\end{equation}
For each code, we use the pair \((Q_i,p_{\mathrm{th},i})\) and fit the data in log-log space as
\begin{equation}
 \log p_{\mathrm{th},i}
 =
 \log C-\alpha\log Q_i.
\end{equation}
Thus, \(\log C\) and \(-\alpha\) are obtained from the intercept and slope of the best-fit line, respectively. The goodness of fit is quantified by
\begin{equation}
 gf\_value
 =
 1-
 \frac{\sum_i (y_i-\hat y_i)^2}
 {\sum_i (y_i-\bar y)^2},
\end{equation}
where \(y_i=\log p_{\mathrm{th}_i}\), \(\hat y_i\) is the fitted value, and \(\bar y\) is the mean of \(y_i\).
Thus, the pseudo-threshold generally decreases as the number of interactions increases.
Figure~\ref{fig:threshold-Q-final} shows this relation for the seven- and eight-qubit records using the bare method.
Values of \(gf\_value\) closer to one indicate better agreement between the data and the fitted relation.
Table~\ref{tab:power-law-final} reports the fitted parameters for the relation \(p_{\mathrm{th}}\simeq C Q^{-\alpha}\).
Since every fitted exponent \(\alpha\) is positive, all fitted models predict that the pseudo-threshold decreases as \(Q\) increases.
Figure~\ref{fig:threshold-Q-final} shows this relation for the seven- and eight-qubit records using the bare method.
The coefficient of determination \(gf\_value\) measures how well this power law i.e., Eq.\eqref{eq:power-law-final} explains the variation in the logarithm of the pseudo-threshold.
Values of \(gf\_value\) closer to one indicate better agreement between the data and the fitted relation.
Table~\ref{tab:power-law-final} reports the fitted parameters for the relation \(p_{\mathrm{th}}\simeq C Q^{-\alpha}\).
Since every fitted exponent \(\alpha\) is positive, all fitted models predict that the pseudo-threshold decreases as \(Q\) increases.
% Drop-in PGFPlots replacement for the two threshold-versus-Q PDF panels.
% The package contains all 1500 record-level data points for both noise models.
%
% Add the following lines to the manuscript preamble:
%   \usepackage{pgfplots}
%   \pgfplotsset{compat=1.18}
%
% Keep the data directory next to the main manuscript file. If the data are
% stored elsewhere, define \thresholdQDataPath before inputting this file.
\providecommand{\thresholdQDataPath}{.}

\definecolor{sevenRecords}{HTML}{1F77B4}
\definecolor{sevenFit}{HTML}{FF7F0E}
\definecolor{eightRecords}{HTML}{2CA02C}
\definecolor{eightFit}{HTML}{D62728}

\begin{figure}[ht]
\centering

\begin{tikzpicture}[scale=0.95]
\begin{semilogyaxis}[
  width=0.95\columnwidth,
  height=0.64\columnwidth,
  xmin=23.5,
  xmax=51,
  ymin=2.7e-4,
  ymax=1.72e-3,
  xtick={25,30,35,40,45,50},
  ytick={3e-4,4e-4,6e-4,1e-3},
  yticklabels={$3\times10^{-4}$,$4\times10^{-4}$,
    $6\times10^{-4}$,$10^{-3}$},
  xlabel={Effective interaction count $Q$},
  ylabel={Pseudo-threshold $p_{\mathrm{th}}$},
  title={Bare code for anisotropic noise},
  tick label style={font=\scriptsize},
  label style={font=\small},
  title style={font=\small},
  grid=both,
  major grid style={draw=gray!28,line width=0.25pt},
  minor grid style={draw=gray!14,line width=0.20pt},
  legend style={font=\tiny,at={(0.98,0.98)},anchor=north east,
    draw=gray!50,fill=white,fill opacity=0.92,text opacity=1,
    cells={anchor=west},
    inner xsep=2pt,
  inner ysep=1pt,
  row sep=-1pt
    },
  legend cell align=left
]
\addplot[
  only marks,
  color=sevenRecords,
  mark=*,
  mark size=1.05pt,
  draw opacity=0.28,
  fill opacity=0.28
] table[x=Q,y=ani] {threshold_vs_Q_713.dat};
\addlegendentry{$[[7,1,3]]$ records}

\addplot[
  color=sevenFit,
  line width=1.25pt,
  domain=24:47,
  samples=250
] {2.660312611806788*x^(-2.34914505690382)};
\addlegendentry{$[[7,1,3]]$ fit}

\addplot[
  only marks,
  color=eightRecords,
  mark=square*,
  mark size=1.05pt,
  draw opacity=0.28,
  fill opacity=0.28
] table[x=Q,y=ani] {threshold_vs_Q_813.dat};
\addlegendentry{$[[8,1,3]]$ records}

\addplot[
  color=eightFit,
  line width=1.25pt,
  domain=25:50,
  samples=250
] {1.177502939431063*x^(-2.057338038687773)};
\addlegendentry{$[[8,1,3]]$ fit}
\end{semilogyaxis}
\end{tikzpicture}

\vspace{0.30cm}

\begin{tikzpicture}[scale=0.95]
\begin{semilogyaxis}[
  width=0.95\columnwidth,
  height=0.64\columnwidth,
  xmin=23.5,
  xmax=51,
  ymin=6.4e-4,
  ymax=4.2e-3,
  xtick={25,30,35,40,45,50},
  ytick={1e-3,2e-3,3e-3,4e-3},
  yticklabels={$10^{-3}$,$2\times10^{-3}$,
    $3\times10^{-3}$,$4\times10^{-3}$},
  xlabel={Effective interaction count $Q$},
  ylabel={Pseudo-threshold $p_{\mathrm{th}}$},
  title={Bare code for depolarizing noise},
  tick label style={font=\scriptsize},
  label style={font=\small},
  title style={font=\small},
  grid=both,
  major grid style={draw=gray!28,line width=0.25pt},
  minor grid style={draw=gray!14,line width=0.20pt},
  legend style={font=\tiny,at={(0.98,0.98)},anchor=north east,
    draw=gray!50,fill=white,fill opacity=0.92,text opacity=1,
    cells={anchor=west},
    inner xsep=2pt,
  inner ysep=1pt,
  row sep=-1pt},
  legend cell align=left
]
\addplot[
  only marks,
  color=sevenRecords,
  mark=*,
  mark size=1.05pt,
  draw opacity=0.28,
  fill opacity=0.28
] table[x=Q,y=dep] {threshold_vs_Q_713.dat};
\addlegendentry{$[[7,1,3]]$ records}

\addplot[
  color=sevenFit,
  line width=1.25pt,
  domain=24:47,
  samples=250
] {6.165621195398746*x^(-2.3354989185037267)};
\addlegendentry{$[[7,1,3]]$ fit}

\addplot[
  only marks,
  color=eightRecords,
  mark=square*,
  mark size=1.05pt,
  draw opacity=0.28,
  fill opacity=0.28
] table[x=Q,y=dep] {threshold_vs_Q_813.dat};
\addlegendentry{$[[8,1,3]]$ records}

\addplot[
  color=eightFit,
  line width=1.25pt,
  domain=25:50,
  samples=250
] {7.365358380093986*x^(-2.345933311950338)};
\addlegendentry{$[[8,1,3]]$ fit}
\end{semilogyaxis}
\end{tikzpicture}

\caption{Bare-method pseudo-threshold as a function of the effective interaction count \(Q\) for the \( [[7,1,3]] \) and \( [[8,1,3]] \) records.
The upper and lower panels show anisotropic and depolarizing noise, respectively.
The solid curves are the separate fits $p_{\mathrm{th}}=C Q^{-\alpha}$ for each code length.}
\label{fig:threshold-Q-final}
\end{figure}

\begin{table}[ht]
\color{black}
\scriptsize
\setlength{\tabcolsep}{2.1pt}
\renewcommand{\arraystretch}{1.10}
\begin{ruledtabular}
\begin{tabular}{@{}cllccc@{}}
$n$ & Method & Noise & $C$ & $\alpha$ $(95\%\ \mathrm{CI})$ & $gf\_value$ \\
\hline
7 & Bare & Aniso. & 2.660 & 2.349 (2.296, 2.403) & 0.917 \\
8 & Bare & Aniso. & 1.178 & 2.057 (2.004, 2.111) & 0.842 \\
7 & Bare & Depol. & 6.166 & 2.335 (2.288, 2.383) & 0.949 \\
8 & Bare & Depol. & 7.365 & 2.346 (2.304, 2.388) & 0.924 \\
7 & Flag & Aniso. & 2.193 & 2.312 (2.255, 2.369) & 0.910 \\
8 & Flag & Aniso. & 0.883 & 1.992 (1.938, 2.047) & 0.824 \\
7 & Flag & Depol. & 2.562 & 2.131 (2.085, 2.176) & 0.945 \\
8 & Flag & Depol. & 3.217 & 2.160 (2.121, 2.199) & 0.920 \\
\end{tabular}
\end{ruledtabular}
\caption{Power-law fits of the pseudo-threshold as a function of the number of interactions \(Q\) as mentioned in Eq.~\eqref{eq:power-law-final}.
The table reports the fitted parameters, their \(95\%\) confidence intervals, and the quality of the fit measured by \(gf_{value}\).}
\label{tab:power-law-final}
\end{table}

\noindent Under depolarizing noise, the value of $\alpha's$ for the seven- and eight-qubit codes are \(2.335\) and \(2.346\) for the bare method; \(2.131\) and \(2.160\) for flag method.
Thus, increasing the number of interactions gives a similar reduction in pseudo-threshold for both code lengths.
Under anisotropic noise, the seven-qubit codes have larger exponents than the eight-qubit codes.
The bare exponents are \(2.349\) and \(2.057\), while the flag exponents are \(2.312\) and \(1.992\), respectively for $[[7,1,3]]$ graph code.
Therefore, under anisotropic noise, the pseudo-threshold decreases more rapidly with \(Q\) for the seven-qubit codes.

\noindent The values of \(gf\_value\), ranging from \(0.824\) to \(0.949\), show that the power-law relation describes the data well.
However, \(gf\_value\) is consistently lower for the eight-qubit codes than for the corresponding seven-qubit codes.
This suggests that the number of interactions accounts for most of the variation in the threshold, but additional circuit and code-structure properties also influence the eight-qubit results.

\subsection{Comparison between bare and flag pseudo-thresholds}
\label{subsec: comparision between bare-flag-final}
\noindent We compare the bare and flag pseudo-thresholds for each code record under the same noise model.
Table~\ref{tab:bare-flag-final} summarizes the median thresholds and the paired record-level ratios.
For seven-qubit codes, bare extraction gives a higher pseudo-threshold in $450$ of $500$ records under anisotropic noise and in all $500$ cases under depolarizing noise.
At eight-qubit codes, the corresponding counts are $852$ of $1000$ and $999$ of $1000$ for anisotropic and depolarizing noise, respectively.

\begin{table}[ht]
\color{black}
\footnotesize
\setlength{\tabcolsep}{3.0pt}
\renewcommand{\arraystretch}{1.10}
\begin{ruledtabular}
\begin{tabular}{@{}clrrr@{}}
Code & Noise & Bare & Flag & Bare/Flag \\
\hline
$[[7,1,3]]$ & Aniso. & 0.804 & 0.757 & 1.068 \\
$[[7,1,3]]$ & Depol. & 1.901 & 1.605 & 1.183 \\
$[[8,1,3]]$ & Aniso. & 0.664 & 0.629 & 1.054 \\
$[[8,1,3]]$ & Depol. & 1.436 & 1.243 & 1.166 \\
\end{tabular}
\end{ruledtabular}
\caption{Median pseudo-thresholds in units of $10^{-3}$.  The final column
is the median of the paired record-level threshold ratios, not the ratio of
the two displayed medians.}
\label{tab:bare-flag-final}
\end{table}

\noindent Under anisotropic noise, the median bare-to-flag ratio is close to one: $1.068$ at seven qubits and $1.054$ at eight qubits.
The bare advantage is larger under depolarizing noise, with median ratios of $1.183$ and $1.166$, respectively.
Overall, the bare method gives a higher pseudo-threshold than the flag method in most cases, especially under depolarizing noise.
This shows that a carefully selected bare circuit can perform as well as or better than the corresponding flag circuit without using an additional flag qubit.
However, it does not establish a universal ordering between the two extraction methods.

\noindent Taken together, these results suggest a practical procedure for selecting syndrome-extraction circuits.
Among the valid circuits, prefer stabilizer generators and interaction orders with smaller \(Q\), since lower interaction counts are strongly associated with higher pseudo-thresholds.
When several circuits have similar \(Q\), the hook-syndrome count \(N_h\) can be used as a secondary criterion.
Parent-graph sparsity may help guide the search when the canonical stabilizer generators are retained, but it becomes less reliable after generator optimization.

\section{Conclusion}
\label{sec:conclusion}
\noindent This work developed a general framework for finding valid BAC from $[[n,k,3]]$ stabilizer codes. 
Starting with the condition required for hook errors, we gave a search procedure that jointly selects the stabilizer generators and their data-ancilla interaction orders.

\noindent We also established two methods for constructing larger BACs from valid BAC parents.
The pendent extension adds physical qubits while preserving the number of logical qubits and minimum distance $i.e$ $d=3$ in our case.
Blockwise composition combines independent code blocks, thereby increasing the number of logical qubits.
Together, these constructions prove the existence of BACs for larger block lengths under the stated assumptions of the fault model.
Although they do not claim the optimal qubit required to encode that global $k$ logical information.
For the correlated ancilla-data fault model, $94.02\%$ of the enumerated $[[7,1,3]]$ cases and all enumerated eight-qubit cases have a valid BAC.
Many $[[7,1,3]]$ codes require noncanonical stabilizer generators, showing that the stabilizer generators and interaction order are essential parts for the BACs.
In the numerical study, the interaction count $Q$ is the strongest predictor of the pseudo-threshold.
Circuits with smaller $Q$ generally perform better, and the selected bare codes usually match or outperform the corresponding flag method using the bare method, with the clearest advantage in the presence of depolarizing noise.
All the data are publicly available \cite{Gupta2026Dataset}.

\noindent The observed power-law relation between the pseudo-threshold and $Q$ is empirical and depends on the code family, noise model, and decoder.  
The count $Q$ also does not describe the Pauli types or locations of the interactions.
A location-resolved analysis of the faults and analysis on larger codes would help determine which circuit features control logical performance beyond the total interaction count.

\begin{acknowledgments}

\noindent The authors thank Rahul Garg for his assistance with the proofs.
R.J. acknowledges funding support from the Department of Science and Technology Grant No. DST/QTC/NQM/QComm/2024/2. 
H.G. and M.B. thank QuCIS lab for providing the necessary computational resources and IISER Bhopal for the institute fellowship that supports their doctoral work.
The authors used OpenAI ChatGPT (GPT-5.6) and OpenAI Codex tools as interactive aids for exploratory discussion, editing, LaTeX preparation, and auxiliary code/plot checks.
The authors remain fully responsible for this work.
\end{acknowledgments}

\bibliography{apssamp}% Produces the bibliography via BibTeX.
\newpage
\appendix
\color{black}
\section{MBQC Encoding }
\label{APPENDIX: MBQC min distance}
\noindent A graph state is represented by a simple undirected graph $\mathsf G=(\mathcal V_G,\mathcal E_G)$.
The vertices represent qubits, and the edges represent controlled-$Z$ interactions.
For a vertex $i\in\mathcal V_G$, let $\mathcal N(i)$ denote its neighborhood.
The product of all controlled-$Z$ gates incident on $i$ can be written as:
\begin{equation}
    \prod_{(i,j)\in \mathcal E_i}CZ_{ji} = \prod_{j\in \mathcal{N}_i} I_j \otimes \ket{0}\bra{0}_i + \prod_{j\in \mathcal{N}_i}Z_j\otimes \ket{1}\bra{1}_i
    \label{cz equation}
\end{equation}
where $\mathcal E_i$ represents the set of all edges connected to node $i$ and $\mathcal{N}_i$ the set of all neighboring nodes of node $ i \in \mathcal V_G$.

\begin{proof}
Let $|\phi_m\rangle=\alpha|0\rangle_m+\beta|1\rangle_m $ be the state of the message qubit, and initialize the $n$ data qubits in $|+\rangle^{\otimes n}$.
After applying the controlled-$Z$ gates, the joint state is
\begin{equation*}
\begin{split}
    \ket{\psi_m} = \ & \prod_{i \in \{1,2,\cdots,n\}} \prod_{ \substack{(i,j) \in \mathcal E_i \\ j>i }}CZ_{ij} \left ( \ket{+}^{\otimes n}\ket{\phi_{m}}\right ) \\
     = \ &  \Biggl ( \prod_{i \in \{1,2,\cdots,n\}} \prod_{ \substack{(i,j) \in \mathcal E_i\setminus	\mathcal E_m \\  j>i }}CZ_{ij}  \ \biggl( \\  & \prod_{\substack{(m,k)\in \mathcal E_{m} \\ k \in (1,2,\cdots ,n)}}CZ_{mk} \left ( \ket{+}^{\otimes n} \ket{\phi_{m}}\right ) \biggl )\Biggl ) \\
     \ & \text{where $\mathcal E_{m}$ represents the set of all the edges} \\ & \text{connected to the message qubit and using} \\ & \text{Eq.\eqref{cz equation}, we get}\\
     = \ &  \left ( \prod_{i \in \{1,2,\cdots,n\}} \prod_{ \substack{(i,j) \in \mathcal E_i\setminus	\mathcal E_m \\  j>i }}CZ_{ij}\right)  \Biggr [ \biggl (  
    \prod_{k\in \mathcal{N}_{m}} \\ & I_k \otimes \ket{0}\bra{0}_{m}+\prod_{k\in \mathcal{N}_{m}}Z_k \otimes \ket{1}\bra{1}_{m} \biggl )  \ket{+}^{\otimes n}\ket{\phi_{m}}\Biggr ] \\
    = \ &  \left ( \prod_{i \in \{1,2,\cdots,n\}} \prod_{ \substack{(i,j) \in \mathcal E_i\setminus	\mathcal E_m \\  j>i }}CZ_{ij}\right)  \Biggr[  \alpha  \ket{+}^{\otimes n} \ket{0}_{m}  +  \\ & \beta \prod_{k\in \mathcal{N}_{m}}Z_k   \ket{+}^{\otimes n} \ket{1}_{m}\Biggr] \\
    = \ & \alpha \mathrm{U} \ket{+}^{\otimes n }  \ket{0}_{m} +  \beta \mathrm{U}  \prod_{k\in \mathcal{N}_{m}}Z_k \ket{+}^{\otimes n} \ket{1}_{m}, \\ &  \text{where } \mathrm{U}  = \prod_{i \in \{1,2,\cdots,n\}} \prod_{ \substack{(i,j) \in \mathcal E_i\setminus	\mathcal E_m \\  j>i }}CZ_{ij}. \\
    = \ & \alpha \ket{G}  \ket{0}_{m} +  \beta  \prod_{k\in \mathcal{N}_{m}}Z_k \ket{G}\ket{1}_{m}, \\ & \text{where} \ket{G} = \mathrm{U} \ket{+}^{\otimes n }.
\end{split}
\end{equation*}
Hence 
\begin{equation*}
\begin{split}
    \ket{\psi}=\ & \biggl( \alpha \ket{G} + \beta\prod_{k\in \mathcal{N}_{m}}Z_k\ket{G} \biggl) \dfrac{\ket{+}_m}{2} \quad + \\ &\biggl( \alpha \ket{G} - \beta\prod_{k\in \mathcal{N}_{m}}Z_k\ket{G} \biggl) \dfrac{\ket{-}_m}{2}.
\end{split}
\end{equation*}
\noindent After measuring the message qubit in the $X$ basis, the data state is $\alpha|G\rangle + (-1)^{s_m}\beta \prod_{k\in\mathcal N(m)}Z_k|G\rangle,$ where $s_m=0$ for outcome $|+\rangle_m$ and $s_m=1$ for outcome $|-\rangle_m$.
\end{proof}

\subsection{Matrix-based encoding method}
\label{Appendix: matrix method for encoding}
\noindent At each iteration, a pivot row with a ‘1’ in the \((n+i)^{\text{th}}\) column of \(H_Z\) of message qubit is selected, and the pivot is added (modulo 2) to all other rows containing a ‘1’ in that column, thereby isolating the qubit in a single stabilizer.
The pivot row and the two columns corresponding to the measured qubit are subsequently removed. Repeating this procedure for all $k$ message qubits reduces the matrix dimension from $n$ $\times$ $2(n+k)$ to $(n-k)$ $\times$ $2n$, resulting in the stabilizer description of the encoded, generally non-CSS,  $[[n,k,d]]$ code in the MBQC framework.\\
{\linespread{1.05}\selectfont
\begin{algorithm}[H]
\DontPrintSemicolon
\caption{Graph code stabilizer construction via parity check matrix using measurements}
\label{algo: matrix method}

\KwInput{Parity-check matrix \\
\(\ \ \ \ \ \ H= [H_X\mid H_Z]=[I_{n\times n}:0_{n\times k} \mid A^{\mathrm{cc}}:A^{\mathrm{cm}}]\)}
\KwOutput{Updated matrix \(H\) of dimension \((n-k) \times 2n\)}

$\mathtt{m} \gets k$\;
\For{\(\mathtt{iter} \gets 1\) \textbf{to} \(k\)}{
    \(\mathtt{target} \gets 2(n + \mathtt{m})\), \ 
    \(\mathtt{pivot} \gets \mathtt{\emptyset}\)\;
\For{\(\mathtt{i} \gets 1\) \textbf{to} \(\operatorname{row}(H)\)}{
    \If{\(H[\mathtt{i}][\mathtt{target}] = 1\)}{
        $\mathtt{pivot}$ \(\gets \mathtt{i}\)\;
        \textbf{break}
    }
}

\For{\(\mathtt{j} \gets 1\) \textbf{to} \(\operatorname{row}(H)\)}{
    \If{\(\mathtt{j} \neq \mathtt{pivot}\) \textbf{and} \(H[\mathtt{j}][\mathtt{target}] = 1\)}{
        \(H[\mathtt{j}] \gets H[\mathtt{j}] + H[\texttt{pivot}]\) $\bmod 2$
    }
}
$H \gets H[:][\{1, \cdots, 2(n+k)\}\backslash \{n+\mathtt{m},\mathtt{target}\} ]$\\
% Remove columns \((n+k)\) and \(2(n+k)\) from \(\mathbf{H}\)\;
$H \gets H[\{1, \cdots,n\}\backslash \{\texttt{pivot}\}][:]$ \;
$\mathtt{m} \gets \mathtt{m} - 1$
}
\Return{\(H\)}
\end{algorithm}
}

\color{black}
\section{Rate-optimal graph codes}
\label{app:rate-optimal-certificates}

\noindent This section provides explicit code for each fault model, along with an optimized code rate we discovered.
The six-qubit code satisfies the ancilla-only model $\mathcal M_{\mathrm a}$, whereas the seven-qubit code satisfies the correlated ancilla-data model $\mathcal M_{\mathrm c}$.
Figure~\ref{fig:short-block-certificate-graphs} shows the two graph-message constructions.
Blue vertices are data qubits, and the orange vertex is the message qubit measured in the $X$ basis.

\begin{figure}[t]
\centering
\begin{minipage}{0.47\textwidth}
\centering

\begin{tikzpicture}[
  scale=0.55,
  data/.style={circle,draw=blue!70,fill=blue!15,minimum size=7mm,
               inner sep=0pt,font=\bfseries},
  message/.style={circle,draw=black,fill=orange!70,minimum size=7mm,
                  inner sep=0pt,font=\bfseries},
  edge/.style={draw,thick},
  message edge/.style={draw=orange!80!black,very thick,densely dashed}]
\node[data]    (q0) at ( 3.00, 3.00) {$0$};
\node[data]    (q1) at ( 1.40, 0.00) {$1$};
\node[data]    (q2) at ( 3.00,-3.00) {$2$};
\node[data]    (q4) at (-1.40, 0.00) {$4$};
\node[data]    (q5) at (-3.00,-3.00) {$5$};
\node[data]    (q3) at (-3.00, 3.00) {$3$};
\node[message] (q6) at ( 0.00, 1.50) {$6$};
\begin{scope}[on background layer]
\draw[edge] (q0)--(q1);
\draw[edge] (q0)--(q2);
\draw[edge] (q0)--(q3);
\draw[edge] (q1)--(q2);
\draw[edge] (q1)--(q4);
\draw[edge] (q1)--(q5);
\draw[edge] (q2)--(q4);
\draw[edge] (q2)--(q5);
\draw[edge] (q3)--(q5);
\draw[edge] (q4)--(q5);
\draw[message edge] (q6)--(q0);
\draw[message edge] (q6)--(q3);
\draw[message edge] (q6)--(q4);
\end{scope}
\end{tikzpicture}

\smallskip
\small (a) Six data qubits and message vertex $6$.
\end{minipage}
\hfill
\begin{minipage}{0.47\textwidth}
\centering

\begin{tikzpicture}[
  scale=0.53,
  data/.style={circle,draw=blue!70,fill=blue!15,minimum size=6mm,
               inner sep=0pt,font=\small\bfseries},
  message/.style={circle,draw=black,fill=orange!70,minimum size=7mm,
                  inner sep=0pt,font=\small\bfseries},
  edge/.style={draw,thick},
  message edge/.style={draw=orange!80!black,thick,densely dashed,
                       preaction={draw=white,solid,line width=3pt}}]
\node[data]    (q0) at (-3.60, 3.60) {$0$};
\node[data]    (q5) at (-2.00, 0.00) {$5$};
\node[data]    (q6) at (-3.60,-3.60) {$6$};
\node[data]    (q3) at ( 3.60,-3.60) {$3$};
\node[data]    (q1) at ( 0.00,-2.00) {$1$};
\node[data]    (q4) at ( 2.00, 0.00) {$4$};
\node[data]    (q2) at ( 3.60, 3.60) {$2$};
\node[message] (q7) at ( 0.00, 0.20) {$7$};
\begin{scope}[on background layer]
\draw[edge] (q0)--(q2);
\draw[edge] (q0)--(q5);
\draw[edge] (q1)--(q3);
\draw[edge] (q1)--(q4);
\draw[edge] (q2)--(q4);
\draw[edge] (q3)--(q6);
\draw[edge] (q5)--(q6);
\draw[edge] (q0)--(q6);
\draw[edge] (q6)--(q1);
\draw[edge] (q3)--(q2);
\foreach \vertex in {q0,q1,q2,q3,q4,q5,q6}{
  \draw[message edge] (q7)--(\vertex);
}
\end{scope}
\end{tikzpicture}

\smallskip
\small (b) Seven data qubits and message vertex $7$.
\end{minipage}
\caption{Graph-message constructions for the two short-block certificates.
Solid edges join data qubits, and dashed orange edges join the message qubit to its neighbors.
Fig.(a) gives the $[[6,1,3]]$ code used under $\mathcal M_{\mathrm a}$, and Fig.(b) gives the $[[7,1,3]]$ code used under
$\mathcal M_{\mathrm c}$.}
\label{fig:short-block-certificate-graphs}
\end{figure}

\subsection{$[[6,1,3]]$ code for the ancilla-only model}
\label{app:six-qubit-certificate}
\noindent For the graph in Fig.~\ref{fig:short-block-certificate-graphs}(a), message vertex $6$ is measured in $X$- basis gives a $[[6,1,3]]$ code.
Table~\ref{tab:six-qubit-certificate} defines the stabilizer generators of the code with their respective interaction orders.

\begin{table}[ht]

\centering
\begin{ruledtabular}
\begin{tabular}{ccc}
Generator & Pauli string & Interaction order \\
\hline
$g_1$ & $ZZZXXI$ & $Z_2\to X_4\to Z_0\to X_3\to Z_1$ \\
$g_2$ & $IZXYXY$ & $Y_5\to Z_1\to X_4\to Y_3\to X_2$ \\
$g_3$ & $YIIXZY$ & $Y_5\to X_3\to Z_4\to Y_0$ \\
$g_4$ & $IZZZZX$ & $Z_3\to Z_2\to Z_4\to Z_1\to X_5$ \\
$g_5$ & $IYYIII$ & $Y_0\to Y_2$ \\
\end{tabular}
\end{ruledtabular}
\caption{Stabilizer generators of $[[6,1,3]]$ code with their corresponding interaction order.}
\label{tab:six-qubit-certificate}
\end{table}
\noindent The logical operators are
\begin{equation}
 \overline X_6=Z_0X_3Z_5,
 \qquad
 \overline Z_6=Z_0Z_3Z_4.
 \label{eq:six-logical-pair}
\end{equation}
Table~\ref{tab:errors_due_to_faults_6_1_3} contains all hook errors due to fault model $\mathcal M_{\mathrm a}$.
The one thing we need to take care of is that the $Y$ gate is implemented as an amalgamation of $Z$ and $X$ gates.
The hook error generated due to this is always a stabilizer-equivalent single-qubit error in all the stabilizers except $g_2$. 
The hook errors generated in the stabilizer $g_2$ are $Y_5Z_1$, $Y_3X_2$, and $X_3X_2$, each with a unique syndrome. 

\begin{table}[ht]

    \begin{tabular}{|ccc|}
\hline \hline
Generator  & Operators & Syndrome \\
\hline 
$g_1$ & $Z_2X_4$       & $01111$ \\
$g_1$ & $Z_2X_4Z_0$    & $01011$ \\
$g_2$ & $Y_5Z_1$       & $00011$ \\
$g_2$ & $Y_3X_2$    & $00101$ \\
$g_2$ & $X_3X_2$    & $11001$ \\
$g_3$ & $Z_3Z_2$       & $10101$ \\
$g_3$ & $Y_5X_3$       & $01000$ \\
$g_4$ & $Z_3Z_2Z_4$       & $01101$ \\
\hline\hline
\end{tabular}
\caption{Complete list of hook error due to fault model $\mathcal M_{\mathrm a}$ for [[6,1,3]] code with their corresponding syndrome.}
\label{tab:errors_due_to_faults_6_1_3}
\end{table}

\subsection{$[[7,1,3]]$ code for the correlated ancilla-data model}
\label{app:seven-qubit-certificate}
\noindent For the graph in Fig.~\ref{fig:short-block-certificate-graphs}(b), message vertex $7$ is measured in $X$- basis gives a $[[7,1,3]]$ code.
Table~\ref{tab:seven-qubit-certificate} defines the stabilizer generators of the code with their respective interaction orders.

\begin{table}[ht]
\color{black}
\centering
\setlength{\tabcolsep}{3pt}
\begin{ruledtabular}
\begin{tabular}{ccc}
Generator & Pauli string & Interaction order \\
\hline
$g_1$ & $ZZZXIXI$ & $Z_1\to X_3\to Z_0\to X_5\to Z_2$ \\
$g_2$ & $IIIXXIZ$ & $X_3\to X_4\to Z_6$ \\
$g_3$ & $ZXXIIIZ$ & $X_1\to Z_0\to Z_6\to X_2$ \\
$g_4$ & $XZIXIZI$ & $X_0\to X_3\to Z_1\to Z_5$ \\
$g_5$ & $ZXIZZXI$ & $X_1\to Z_0\to Z_3\to Z_4\to X_5$ \\
$g_6$ & $IZXIZZX$ & $X_2\to X_6\to Z_5\to Z_1\to Z_4$ \\
\end{tabular}
\end{ruledtabular}
\caption{Stabilizer generators of $[[7,1,3]]$ code with their corresponding interaction order.}
\label{tab:seven-qubit-certificate}
\end{table}
\noindent Logical operators are
\begin{equation}
 \overline X_7=Z_0X_5Z_6,
 \qquad
 \overline Z_7=Z_3Y_4Y_5.
 \label{eq:seven-logical-pair}
\end{equation}
Table~\ref{tab:errors_due_to_faults_7} contains all hook errors due to fault model $\mathcal M_{\mathrm c}$.
\begin{table}[t]
\centering
\scriptsize
\setlength{\tabcolsep}{2.2pt}
\renewcommand{\arraystretch}{1.05}
\newcommand{\ErrSyn}[2]{%
  \shortstack[c]{%
    \rule{0pt}{2.8ex}$#1$\\[1.5pt]
    $#2$\rule[-1.2ex]{0pt}{2.4ex}}%
}
\begin{ruledtabular}
\color{black}
\begin{tabular}{|cccc|}
\rule{0pt}{2.6ex}Generator & $E=X$ & $E=Y$ & $E=Z$ \\
\hline

$g_1$ 
& \ErrSyn{Z_1X_3}{001000}
& \ErrSyn{Z_1Y_3}{111100}
& \ErrSyn{Z_1Z_3}{111110} \\
\hline
$g_1$ 
& \ErrSyn{X_0Z_1X_3}{100010}
& \ErrSyn{Y_0Z_1X_3}{100110}
& \ErrSyn{Z_0Z_1X_3}{001100} \\
\hline
$g_1$ 
& \ErrSyn{Z_0Z_1X_3X_5}{001001}
& \ErrSyn{Z_0Z_1X_3Y_5}{101011}
& \ErrSyn{Z_0Z_1X_3Z_5}{101110} \\
\hline
\noalign{\vskip 1.5pt}

$g_2$ 
& \ErrSyn{X_3X_4}{000001}
& \ErrSyn{X_3Y_4}{010001}
& \ErrSyn{X_3Z_4}{010010} \\
\hline
\noalign{\vskip 1.5pt}

$g_3$ 
& \ErrSyn{X_0X_1}{001111}
& \ErrSyn{Y_0X_1}{001011}
& \ErrSyn{Z_0X_1}{100001} \\
\hline
$g_3$ 
& \ErrSyn{Z_0X_1X_6}{111001}
& \ErrSyn{Z_0X_1Y_6}{111000}
& \ErrSyn{Z_0X_1Z_6}{100000} \\
\hline
\noalign{\vskip 1.5pt}

$g_4$ 
& \ErrSyn{X_0X_3}{101000}
& \ErrSyn{X_0Y_3}{011100}
& \ErrSyn{X_0Z_3}{011110} \\
\hline
$g_4$ 
& \ErrSyn{X_0X_1X_3}{001101}
& \ErrSyn{X_0Y_1X_3}{000111}
& \ErrSyn{X_0Z_1X_3}{100010} \\
\hline
\noalign{\vskip 1.5pt}

$g_5$ 
& \ErrSyn{X_0X_1}{001111}
& \ErrSyn{Y_0X_1}{001011}
& \ErrSyn{Z_0X_1}{100001} \\
\hline
$g_5$ 
& \ErrSyn{Z_0X_1X_3}{100011}
& \ErrSyn{Z_0X_1Y_3}{010111}
& \ErrSyn{Z_0X_1Z_3}{010101} \\
\hline
$g_5$ 
& \ErrSyn{Z_0X_1Z_3X_4}{010110}
& \ErrSyn{Z_0X_1Z_3Y_4}{000110}
& \ErrSyn{Z_0X_1Z_3Z_4}{000101} \\
\hline
\noalign{\vskip 1.5pt}

$g_6$
& \ErrSyn{X_2X_6}{111000}
& \ErrSyn{X_2Y_6}{111001}
& \ErrSyn{X_2Z_6}{100001} \\
\hline
$g_6$ 
& \ErrSyn{X_2X_5X_6}{111101}
& \ErrSyn{X_2Y_5X_6}{011111}
& \ErrSyn{X_2Z_5X_6}{011010} \\
\hline
$g_6$ 
& \ErrSyn{X_1X_2Z_5X_6}{111111}
& \ErrSyn{Y_1X_2Z_5X_6}{110101}
& \ErrSyn{Z_1X_2Z_5X_6}{010000} \\

\end{tabular}
\end{ruledtabular}
\caption{Complete list of hook error due to fault model $\mathcal M_{\mathrm c}$ for [[7,1,3]] code with their corresponding syndrome.}
\label{tab:errors_due_to_faults_7}
\end{table}
There are ten syndromes that are the same as either another hook error or a single qubit error, but all the pairs are stabilizer equivalent, as listed in Table~\ref{tab:seven-safe-degeneracies}.

\begin{table}[t]
\color{black}
\centering
\begin{ruledtabular}
\begin{tabular}{ccc}
Syndrome & Equal-syndrome pair $\{E,E'\}$ & $E^{\dagger}E'$ \\
\hline
$000001$ & $\{Z_6,\ X_3X_4\}$ & $g_2$ \\
$000101$ & $\{X_5,\ Z_0X_1Z_3Z_4\}$ & $g_5$ \\
$001001$ & $\{Z_2,\ Z_0Z_1X_3X_5\}$ & $g_1$ \\
$010000$ & $\{Z_4,\ Z_1X_2Z_5X_6\}$ & $g_6$ \\
$100000$ & $\{X_2,\ Z_0X_1Z_6\}$ & $g_3$ \\
$100001$ & $\{X_2Z_6,\ Z_0X_1\}$ & $g_3$ \\
$100010$ & $\{Z_5,\ X_0Z_1X_3\}$ & $g_4$ \\
$101110$ & $\{Y_0,\ Z_0Z_1X_3Z_5\}$ & $g_4$ \\
$111000$ & $\{X_2X_6,\ Z_0X_1Y_6\}$ & $g_3$ \\
$111001$ & $\{X_2Y_6,\ Z_0X_1X_6\}$ & $g_3$ \\
\end{tabular}
\end{ruledtabular}
\caption{Complete list of safe equal-syndrome pairs for the seven-qubit
correlated certificate. The last column gives the phase-free product
\(E^{\dagger}E'\).}
\label{tab:seven-safe-degeneracies}
\end{table}

\end{document}